\documentclass[11pt,letterpaper]{article}
\usepackage[utf8]{inputenc} 
\usepackage{amssymb,amsmath,amsthm,graphicx}
\usepackage{natbib}
\usepackage{enumitem} 

\newtheorem{theorem}{Theorem}

\DeclareMathOperator*{\med}{med}
\newcommand{\norm}[1]{\left\lVert#1\right\rVert}

\newcommand{\bzero}{\boldsymbol 0}
\newcommand{\bc}{\boldsymbol c}

\newcommand{\bv}{\boldsymbol v}
\newcommand{\bx}{\boldsymbol x}
\newcommand{\by}{\boldsymbol y}
\newcommand{\bz}{\boldsymbol z}
\newcommand{\bmu}{\boldsymbol \mu}
\newcommand{\btheta}{\boldsymbol \theta}

\newcommand{\bSigma}{\boldsymbol \Sigma}

\begin{document}

\title{Multivariate and functional classification\\ 
       using depth and distance}
\author{Mia Hubert, Peter J. Rousseeuw, Pieter Segaert}
\date{July 7, 2016}

\maketitle

\begin{abstract}
We construct classifiers for multivariate and functional 
data.
Our approach is based on a kind of distance between data
points and classes.
The distance measure needs to be robust to outliers and
invariant to linear transformations of the data.
For this purpose we can use the {\it bagdistance} which 
is based on halfspace depth.
It satisfies most of the properties of a norm but is able 
to reflect asymmetry when the class is skewed.
Alternatively we can compute a measure of outlyingness
based on the skew-adjusted projection depth.
In either case we propose the {\it DistSpace} transform
which maps each data point to the vector of its distances
to all classes, followed by $k$-nearest neighbor (kNN)
classification of the transformed data points.
This combines invariance and robustness with the simplicity 
and wide applicability of kNN.
The proposal is compared with other methods in experiments 
with real and simulated data.\\
\end{abstract}


\section{Introduction} 
\label{sec:introduction}

Supervised classification of multivariate data is a common 
statistical problem. 
One is given a training set of observations and their membership 
to certain groups (classes).
Based on this information, one must assign new observations to
these groups. 
Examples of classification rules include, but are not limited to, 
linear and quadratic discriminant analysis, $k$-nearest neighbors 
(kNN), support vector machines, and decision trees. 
For an overview see e.g. \cite{Hastie:StatLearning}. 

However, real data often contain outlying observations.
Outliers can be caused by recording errors or typing 
mistakes, but they may also be valid observations that 
were sampled from a different population.  
Moreover, in supervised classification some observations in the
training set may have been mislabeled, i.e.\ attributed to the
wrong group. To reduce the potential effect of outliers on the data analysis and to detect them, many robust methods have been developed, see e.g. \cite{Rousseeuw:RobReg} and \cite{Maronna:RobStat}. 

Many of the classical and robust classification methods 
rely on distributional assumptions such as multivariate
 normality or elliptical symmetry \citep{Hubert:Discrim}. 
Most robust approaches that can deal with more general
data make use of the concept of depth, which measures the
centrality of a point relative to a multivariate sample.
The first type of depth was the halfspace depth 
of \cite{Tukey:depth}, followed by other depth functions 
such as simplicial depth \citep{Liu:Simplicial} and 
projection depth \citep{Zuo:Depth}.

Several authors have used depth in the context of classification.
\cite{Christmann:Overlap} and \cite{Christmann:RegrDeptSVM} 
applied regression depth \citep{Rousseeuw:Regdepth}.
The maximum depth classification rule of \cite{Liu:Simplicial}
was studied by \cite{Ghosh:Maxdepth}
and extended by \cite{Li:DDplot}. 
\cite{Dutta:depth} used projection depth.

In this paper we will present a novel technique called 
{\it classification in distance space}. 
It aims to provide a fully non-parametric tool for the robust 
supervised classification of possibly skewed multivariate data. 
In Sections \ref{sec:MultivariateDepthAndDistanceMeasures} and 
\ref{sec:SkewAdjustedProjectionDepth} we will describe the key 
concepts needed for our construction.
Section \ref{sec:MultivariateClassifiers} discusses some 
existing multivariate classifiers and introduces our approach. 
A thorough simulation study for multivariate data is performed in 
Section \ref{sec:Simulation}. 
From Section \ref{sec:FunctionalDataAndItsTools} onwards 
we will focus our attention on the increasingly important 
framework of {\it functional} data, the analysis of which 
is a rapidly growing field. 
We will start by a general description, and then extend
our work on multivariate classifiers to functional 
classifiers.


\section{Multivariate depth and distance measures}
\label{sec:MultivariateDepthAndDistanceMeasures}
\subsection{Halfspace depth}
\label{sec:HalfspaceDepth}

If $Y$ is a random variable on $\mathbb{R}^p$ with distribution 
$P_{\,Y}$, then the halfspace depth of any point 
$\bx \in \mathbb{R}^p$ relative to $P_{\,Y}$ is defined 
as the minimal probability mass contained in a closed halfspace 
with boundary through $\bx$: 
\begin{equation*}\label{eq:hd}
   \mbox{HD}(\bx;P_{\,Y})=
   \inf_{||\bv||=1} \; P_{\,Y}
	 \left\{ \bv'Y \geqslant  \bv'\bx \right\}.
\end{equation*}
Halfspace depth satisfies the requirements of a statistical
depth function as formulated by \cite{Zuo:Depth}: it is affine invariant (i.e.\ invariant to translations and
nonsingular linear transformations), it attains its maximum
value at the center of symmetry if there is one, it is
monotone decreasing along rays emanating from the center, 
and it vanishes at infinity. 

For any statistical depth function $D$ and for any 
$\alpha \in [0,1]$ the $\alpha$-{\it depth region} $D_\alpha$ 
is the set of points whose depth is at least $\alpha$:
\begin{equation} \label{eq:depthregions}
    D_\alpha= \{ \bx \in \mathbb{R}^p \; ;
		\; \mbox{D}(\bx;P_{\,Y}) \geqslant \alpha \} \; .
\end{equation}
The boundary of $D_\alpha$ is known as 
the $\alpha$-{\it depth contour}. The halfspace depth regions are closed, convex, and nested for 
increasing $\alpha$. Several properties of the halfspace depth function and its contours were studied in 
\cite{Masse:Trimming} and \cite{Rousseeuw:DepthPop}. 
The halfspace median (or Tukey median) is defined as the center 
of gravity of the smallest non-empty depth region, i.e.\ the 
region containing the points with maximal halfspace depth.

The finite-sample definitions of the halfspace depth, the Tukey 
median and the depth regions are obtained by replacing 
$P_{\,Y}$ by the empirical probability distribution $P_n$. 
Many finite-sample properties, including the breakdown value of the Tukey median, were derived in 
\cite{Donoho:Depth}.

To compute the halfspace depth, several affine invariant 
algorithms have been developed.
\cite{Rousseeuw:Bivlocdepth} and \cite{Rousseeuw:Ldepth} provided
exact algorithms in two and three dimensions and an approximate 
algorithm in higher dimensions. Recently \cite{Dyckerhoff:HSdepth} 
developed exact algorithms in higher dimensions.  
Algorithms to compute the halfspace median have been developed 
by \cite{Rousseeuw:Tukeymed} and \cite{Struyf:DeepestLocation}. 
To compute the depth contours the algorithm of \cite{Ruts:isodepth} 
can be used in the bivariate setting, whereas the algorithms 
constructed by \cite{Hallin:Quantiles} and 
\cite{Paindaveine:CompQuantiles} are applicable to at least $p=5$.


\subsection{The bagplot}  
\label{sec:TheBagplot}

\begin{figure}[!ht]
\centering
\includegraphics[width=0.45\textwidth]{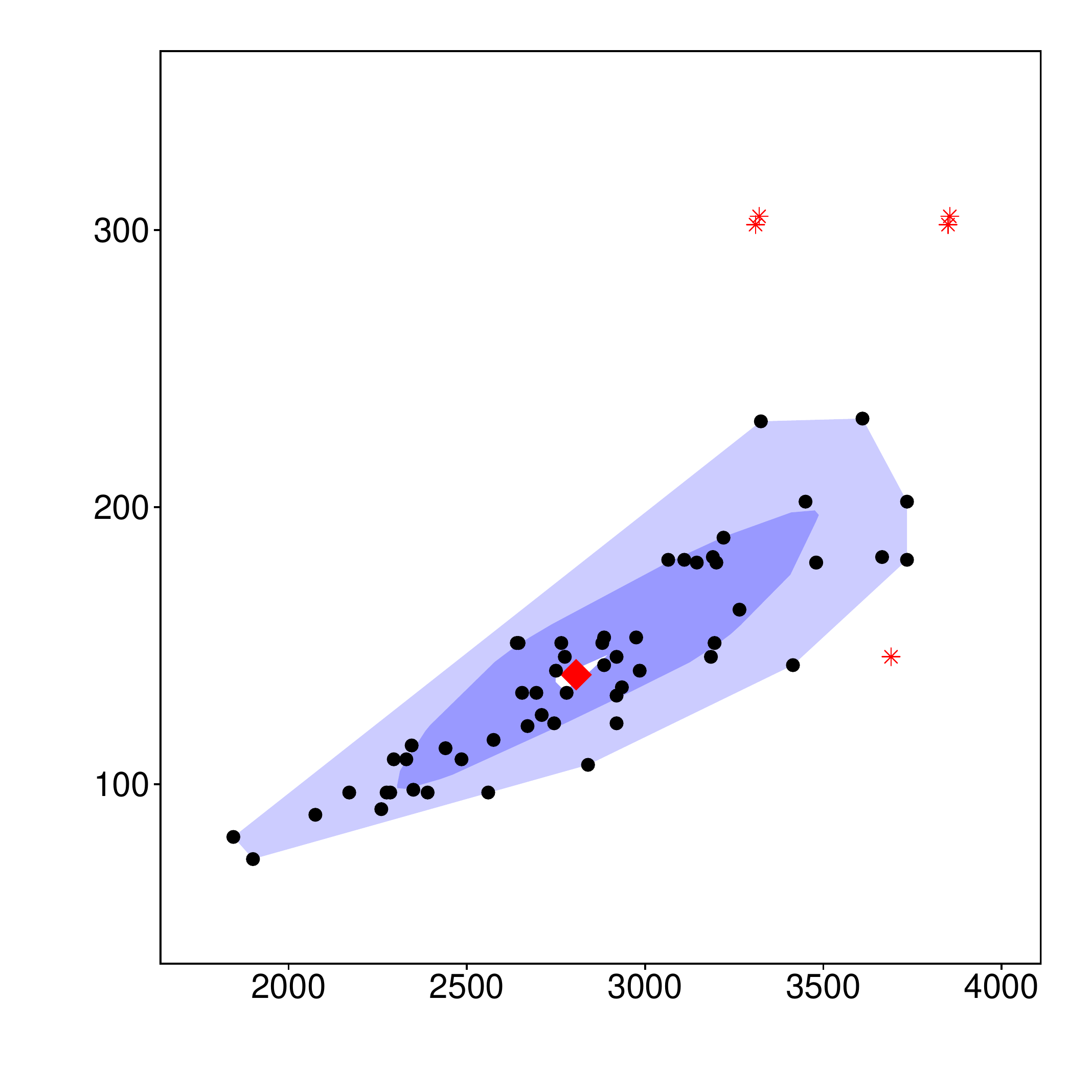}
\vspace{-0.2cm}
\caption{Bagplot of a bivariate dataset.}
\label{fig:bagplot}
\end{figure}

The bagplot of \cite{Rousseeuw:Bagplot} generalizes the 
univariate boxplot to bivariate data, as illustrated in
Figure \ref{fig:bagplot}. 
The dark-colored {\it bag} is the smallest depth region with at 
least 50\% probability mass, i.e.\ $B = D_{\tilde{\alpha}}$ such
that $P_{\,Y}(B) \geqslant 0.5$ and $P_{\,Y}(D_\alpha) < 0.5$ 
for all $\alpha > \tilde{\alpha}$. 
The white region inside the bag is the smallest depth region, 
which contains the halfspace median (plotted as a red diamond). 
The {\it fence}, which itself is rarely drawn, is obtained 
by inflating the bag by a factor 3 relative to the median, and the data points outside of it are flagged as outliers 
and plotted as stars. 
The light-colored {\it loop} is the convex hull of the data points 
inside the fence. 

The bagplot exposes several features of the bivariate data distribution: its center (by the Tukey median), its dispersion 
and shape (through the sizes and shape of the bag and the loop) 
and the presence or absence of  outliers.  
In Figure \ref{fig:bagplot} we see a moderate deviation from 
symmetry as well as several observations that lie outside 
the fence. 
One could extend the notion of bagplot to higher dimensions 
as well, but a graphical representation then becomes harder 
or impossible.

\subsection{The bagdistance} \label{sec:Bagdistance} 
Although the halfspace depth is small in outliers, it does 
not tell us how distant they are from the center of the data. 
Also note that any point outside the convex hull of the data 
has zero halfspace depth, which is not so informative.  
Based on the concept of halfspace depth, we can however derive 
a statistical distance of a multivariate point 
$\bx \in \mathbb{R}^p$ to $P_{\,Y}$ as in~\citep{Hubert:MFOD}. 
This distance uses both the center and the dispersion of 
$P_{\,Y}$. 
To account for the dispersion it uses the bag $B$ defined above. 
Next, $\bc(\bx):=\bc_{\bx}$ is defined as the intersection of 
the boundary of $B$ and the ray from the halfspace 
median $\btheta$ through $\bx$.
The {\it bagdistance} of $\bx$ to $Y$ is then given by the 
ratio of the Euclidean distance of $\bx$ to $\btheta$
and the Euclidean distance of $\bc_{\bx}$ to $\btheta$: 
\begin{equation}
\label{eq:bagdistanceDef}
  bd(\bx;P_{\,Y}) = \left\{
    \begin{array}{ll}
        0  & \,\,\, \mbox{  if } \bx = \btheta \\
        \norm{\bx-\btheta}/\norm{\bc_{\bx}-\btheta}
           & \,\,\, \mbox{  elsewhere} \, .
    \end{array}
  \right.
\end{equation}
The denominator in \eqref{eq:bagdistanceDef} accounts for the 
dispersion of $P_Y$ in the direction of $\bx$. 
Note that the bagdistance does not assume symmetry and is
affine invariant.

\begin{figure}[!ht]
\centering
\includegraphics[width=0.45\textwidth]{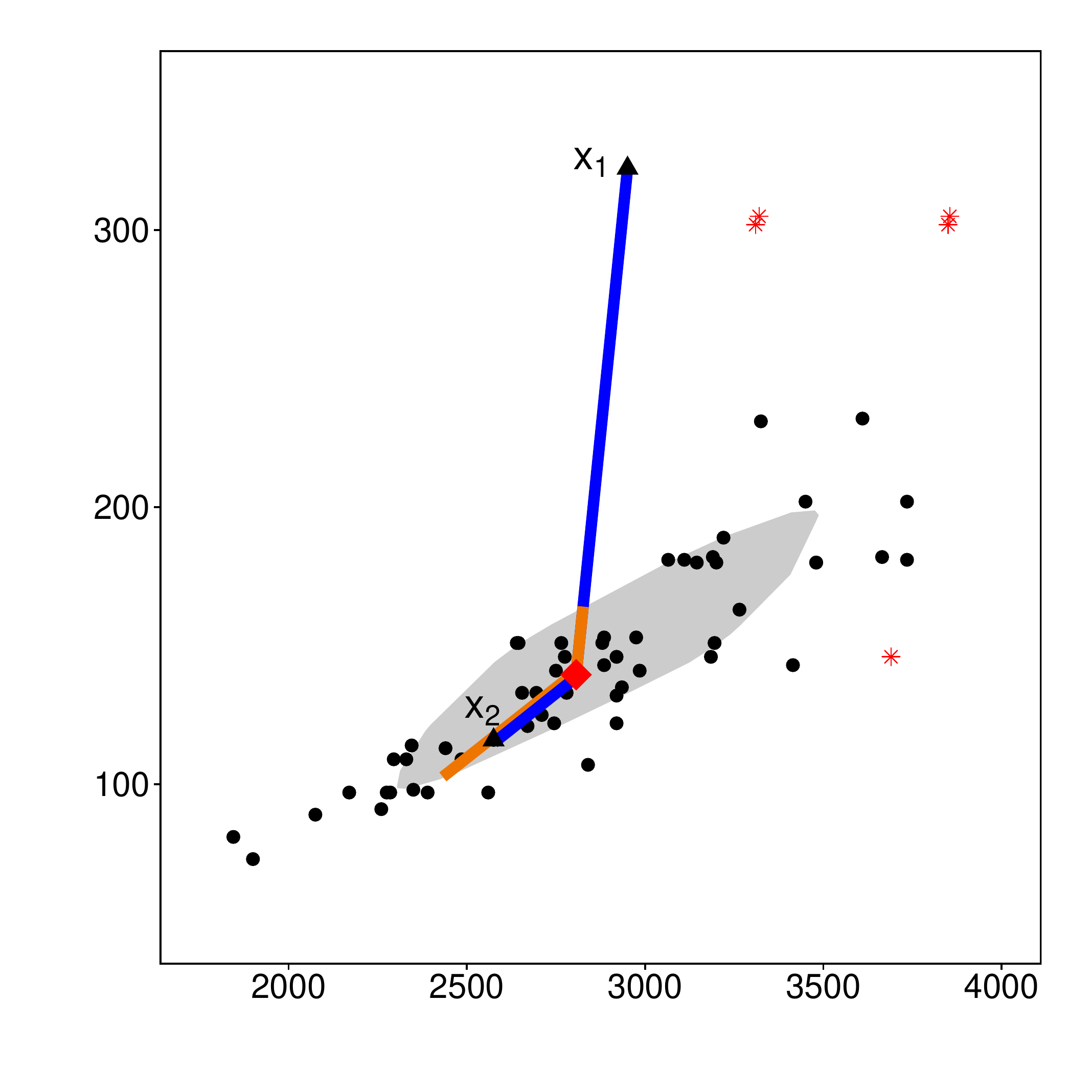}
\vspace{-0.2cm}
\caption{Illustration of the bagdistance between an
         arbitrary point and a sample.}
\label{fig:Bagdistance}
\end{figure}

The finite-sample definition is similar and illustrated 
in Figure~\ref{fig:Bagdistance} for the data set in 
Figure~\ref{fig:bagplot}. 
Now the bag is shown in gray. 
For two new points $\bx_1$ and $\bx_2$ their Euclidean distance 
to the halfspace median is marked by dark blue lines, 
whereas the orange lines correspond to the denominator of 
\eqref{eq:bagdistanceDef} and reflect how these distances will 
be scaled. 
Here the lengths of the blue lines are the same (although they 
look different as the scales of the coordinates axes are quite distinct). 
On the other hand the bagdistance of $\bx_1$ is 7.43 and 
that of $\bx_2$ is only 0.62.
These values reflect the position of the points relative to 
the sample, one lying quite far from the most central half of 
the data and the other one lying well within the central half. 

%

Similarly we can compute the bagdistance of the outliers. 
For the uppermost right outlier with coordinates $(3855,305)$ 
we obtain 4.21, whereas the bagdistance of the lower outlier 
$(3690,146)$ is 3.18. 
Both distances are larger than 3, the bagdistance 
of all points on the fence, but the bagdistance now reflects 
the fact that the lower outlier is merely a boundary case. 
The upper outlier is more distant, but still not as remote 
as $\bx_1$.    

We will now provide some properties of the bagdistance.
We define a generalized norm as a function 
$g: \mathbb{R}^p \to [0, \infty[$ such that $g(\bzero)=0$ 
and $g(\bx) \neq 0$ for $\bx \neq \bzero$, which 
satisfies $g(\gamma \bx)=\gamma g(\bx)$ for all $\bx$ and 
all $\gamma>0$. 
In particular, for a positive definite $p \times p$ 
matrix $\bSigma$ it holds that 
\begin{equation} 
\label{eq:GaussianDistancefunction}  
   g(\bx) = \sqrt{\bx'\bSigma^{-1}\bx} 
\end{equation} 
is a generalized norm (and even a norm). 

Now suppose we have a compact set $B$ which is star-shaped 
about zero, i.e.\ for all $\bx \in B$ 
and $0 \leqslant \gamma \leqslant 1$ it holds 
that $\gamma \bx \in B$. 
For every $\bx \neq \bzero$ we then construct the point 
$\bc_{\bx}$ as the intersection between the boundary of $B$ and 
the ray emanating from $\bzero$ in the direction of $\bx$.
Let us assume that $\bzero$ is in the interior of $B$, that is, 
there exists $\varepsilon>0$ such that the ball 
$B(\bzero,\varepsilon) \subset B$. 
Then $\norm{\bc_{\bx}}>0$ whenever $\bx \neq \bzero$. Now define
\begin{equation}
\label{eq:GenBagDist}
  g(\bx) = \begin{cases}
  0& \text{if} \,\, \bx=\bzero \\
  \frac{\norm{\bx}}{\norm{\bc_{\bx}}} &\text{otherwise}.
  \end{cases}
\end{equation}
Note that we do not really need the Euclidean norm, as we
can equivalently define $g(\bx)$ as\; 
$\inf\{\gamma > 0;\;\gamma^{-1} \bx \in B\}$.
We can verify that $g(\cdot)$ is a generalized norm, 
which need not be a continuous function. 
The following result shows more.

\begin{theorem}
\label{theo:gconvex}
If the set $B$ is convex and compact and 
$\bzero \in \mbox{int}{(B)}$ then the function $g$ defined 
in \eqref{eq:GenBagDist} is a convex function and hence 
continuous. 
\end{theorem}
{\it Proof.} We need to show that
\begin{equation} \label{eq:inequalg}
  g(\lambda \bx + (1-\lambda)\by) \leqslant \lambda g(\bx)
	  + (1-\lambda)g(\by)
		\end{equation}
for any $\bx,\by \in \mathbb{R}^p$ 
and $0 \leqslant \lambda \leqslant 1$. 
In case $\left\{\bzero,\bx,\by \right\}$ are collinear the 
function $g$ restricted to this line is 0 in the origin and goes 
up linearly in both directions (possibly with different slopes) 
so \eqref{eq:inequalg} is satisfied for those $\bx$ and $\by$. 
If $\left\{\bzero,\bx,\by \right\}$ are not collinear they form 
a triangle. 
Note that we can write $\bx = g(\bx)\bc_{\bx}$ and 
$\by = g(\by)\bc_{\by}$ and we will denote
$\bz := \lambda \bx + (1-\lambda)\by$.
We can verify that 
$\widetilde{\bz} := 
(\lambda g(\bx) + (1-\lambda)g(\by))^{-1} \bz$
is a convex combination of $\bc_{\bx}$ and $\bc_{\by}$.
By compactness of $B$ we know that $\bc_{\bx},\bc_{\by} \in B$,
and from convexity of $B$ it then follows that 
$\widetilde{\bz} \in B$.
Therefore
  $$\norm{\bc_{\bz}} = \norm{\bc_{\widetilde{\bz}}}
	  \geqslant \norm{\widetilde{\bz}}$$ 
so that finally		
  $$g(\bz) = \frac{\norm{\bz}}{\norm{\bc_{\bz}}}
    \leqslant \frac{\norm{\bz}}{\norm{\widetilde{\bz}}}
    = \lambda g(\bx) + (1-\lambda)g(\by)\,\,.\quad \square$$ 
Note that this result generalizes Theorem 2 of \cite{Hubert:MFOD} from halfspace depth to general convex sets.
It follows that $g$ satisfies the triangle inequality since 
   $$g(\bx+\by) = 2 g(\frac12 \bx +\frac12 \by) \leqslant
   2\frac12 g(\bx) + 2\frac12 g(\by) = g(\bx)+g(\by) \,\,\, .$$
Therefore $g$ (and thus the bagdistance) satisfies the 
conditions
\begin{itemize}[noitemsep,topsep=0pt]
\item[(i)] $g(\bx) \geqslant 0$ for all $\bx \in \mathbb{R}^p$
\item[(ii)] $g(\bx) = 0$ implies $\bx = \bzero$
\item[(iii)] $g(\gamma \bx) = \gamma g(\bx)$ for all
             $\bx \in \mathbb{R}^p$ and $\gamma \geqslant 0$
\item[(iv)] $g(\bx+\by) \leqslant g(\bx) + g(\by)$ for
            all $\bx,\by \in \mathbb{R}^p$\,\,. 
\end{itemize}
This is almost a norm, in fact, it would become a norm if we 
were to add
$$g(-\bx) = g(\bx) \mbox{ for all } \bx \in \mathbb{R}^p \,\,\,.$$ 
The generalization makes it possible for $g$ to 
reflect asymmetric dispersion. 
(We could easily turn it into a norm by computing
$h(\bx) = (g(\bx) + g(-\bx))/2$ but then we would lose that 
ability.)

Also note that the function $g$ defined in \eqref{eq:GenBagDist} 
does generalize the Mahalanobis distance 
in \eqref{eq:GaussianDistancefunction}, as can be seen by taking 
$B = \left\{\bx; \,\, \bx'\bSigma^{-1}\bx \leqslant 1 \right\}$ 
which implies $\bc_{\bx} = (\bx'\bSigma^{-1}\bx)^{-1/2} \bx$ 
for all $\bx \neq \bzero$ so 
$g(\bx)
  = \norm{\bx}/((\bx'\bSigma^{-1}\bx)^{-1/2}\norm{\bx})
  = \sqrt{ \bx'\bSigma^{-1}\bx} \,\,\,.$

Finally note that Theorem \ref{theo:gconvex} holds whenever 
$B$ is a convex set. 
Instead of halfspace depth we could also use regions of
projection depth or the depth function in 
Section~\ref{sec:SkewAdjustedProjectionDepth}.
On the other hand, if we wanted to describe nonconvex
data regions we would have to switch to a different
star-shaped set $B$ in~\eqref{eq:GenBagDist}.

In the univariate case, the compact convex set $B$ in Theorem
\ref{theo:gconvex} becomes a closed interval which we can denote 
by $B=[-\frac{1}{b},\frac{1}{a}]$ with $a, b > 0$, so that
                $$ g(x) = ax^+ + bx^-\,\,. $$
In linear regression the minimization of 
$\sum_{i=1}^{n} g(r_i)$ yields the $a/(a+b)$ regression
quantile of \cite{Koenker:RegQuant}.

It is straightforward to extend Theorem \ref{theo:gconvex} to a 
nonzero center by subtracting the center first. 

To compute the bagdistance of a point $\bx$ with respect to 
a $p$-variate sample we can first compute the bag and then the 
intersection point $\bc_{\bx}$. 
In low dimensions computing the bag is feasible, and it is worth 
the effort if the bagdistance needs to be computed for many points. 
In higher dimensions computing the bag is harder, and then a 
simpler and faster algorithm is to search for the multivariate 
point $\bc^*$ on the ray from $\btheta$ through $\bx$ such that 
\begin{equation}\label{eq:computebd}
  \text{HD}(\bc^*;P_n) = \med_i \; \{ \text{HD}(\by_i;P_n)\} 
\end{equation}
where $\by_i$ are the data points.
Since HD is monotone decreasing on the ray this can be done 
fairly fast, e.g.\ by means of the bisection algorithm.

Table \ref{table:compbd} lists
the computation time needed to calculate the bagdistance 
of $m \in \left\{1, 50, 100, 1000\right\}$ points with 
respect to a sample of $n=100$ points 
in dimensions $p \in \left\{2, 3, 4, 5\right\}$. For $p=2$ the algorithm of \cite{Ruts:isodepth} is used and~\eqref{eq:computebd} otherwise.
The times are averages over 1000 randomly generated data 
sets. 
In each of the 1000 runs the points were generated from
a centered multivariate normal distribution with a 
randomly generated covariance matrix. 
Note that the time for $m=1$ is essentially that of the
right hand side of~\eqref{eq:computebd}.

\begin{table}[htbp]
\centering
\caption{Computation times for the bagdistance ($n=100$), 
         in units of 0.001 seconds.}
\label{table:compbd}
\vspace{0.2cm}
\begin{tabular}{|r|rrrr|}
\hline
  & \multicolumn{4}{c|}{$m$} \\
  $p$ & 1     & 50   & 100   & 1000  \\ 
\hline
2 & 15.6 &  16.2 & 17.4  & 17.1 \\
3 & 34.8 &  67.8 & 84.1  & 310.2 \\
4 & 45.3 &  88.3 & 107.9 & 377.3 \\
5 & 56.4 & 106.3 & 128.2 & 432.8 \\
\hline      
\end{tabular}
\end{table}


\section{Skew-adjusted projection depth} 
\label{sec:SkewAdjustedProjectionDepth} 
Since the introduction of halfspace depth various other affine 
invariant depth functions have been defined (for an overview
see e.g.~\cite{ Mosler:DepthStat}), among which projection 
depth \citep{Zuo:ProjDepth} which is essentially the 
inverse of the Stahel-Donoho outlyingness (SDO).
The population SDO \citep{Stahel:SDEst, Donoho:SDest} of an arbitrary point $\bx$ with respect to 
a random variable $Y$ with distribution $P_{\,Y}$ is defined as 
\begin{equation*}\label{SDO}
  \mbox{SDO}(\bx;P_{\,Y})=\sup_{||\bv||=1} \;\;  
  \frac{| \; \bv' \bx - \med(\bv'Y) \; |}{\text{MAD}(\bv'Y)}
\end{equation*} 
from which the projection depth is derived:
\begin{equation*}
\label{eq:projdepth}
   \mbox{PD}(\bx;P_{\,Y})=\frac{1}{1+\mbox{SDO}(\bx;P_{\,Y})}\; .
\end{equation*} 

Since the SDO has an absolute deviation in the numerator and 
uses the MAD in its denominator it is best suited for symmetric distributions. 
For asymmetric distributions \cite{Brys:RobICA} proposed the 
adjusted outlyingness (AO) in the context of robust independent 
component analysis. 
It is defined as
$$\mbox{AO}(\bx;P_{\,Y})=
  \sup_{||\bv||=1} \; \text{AO}_1(\bv'\bx;P_{\,\bv'Y})$$
where the univariate adjusted outlyingness $\text{AO}_1$ is given by
\begin{equation}
\label{eq:adjOut}
  \text{AO}_1(z;P_Z) = \left\{
	\begin{array}{ll}
		\frac{z-\med(Z)}{w_2(Z) - \med(Z)}
		      & \;\; \mbox{ if } z > \med(Z) \\
		\frac{\med(Z) - z}{\med(Z) -  w_1(z)}
		      & \;\; \mbox{ if } z \leqslant \med(Z) \; . \\
	\end{array} 
  \right.
\end{equation}
Here
\begin{align*}
	w_1(Z) &= Q_1(Z)-1.5 \, e^{-4\text{MC}(Z)} \, \mbox{IQR}(Z)\\
  w_2(Z) &= Q_3(Z)+1.5 \, e^{+3\text{MC}(Z)} \, \mbox{IQR}(Z)   
\end{align*}
if $\text{MC}(Z) \geqslant 0 $, where $Q_1(Z)$ and $Q_3(Z)$ 
denote the first and third quartile of $Z$, $\text{IQR}(Z) = Q_3(Z)-Q_1(Z)$ and $\text{MC}(Z)$ is robust measure of skewness \citep{Brys:RobSkew}.
If $\text{MC}(Z) < 0$ we replace $(z,Z)$ by $(-z,-Z)$. 
The denominator of \eqref{eq:adjOut} corresponds to the 
fence of the univariate adjusted boxplot proposed 
by \cite{Hubert:AdjBoxplot}.

The {\it skew-adjusted projection depth} (SPD) is then given by
\citep{Hubert:MFOD}:
\begin{equation*}\label{eq:spd}
\mbox{SPD}(\bx;P_{\,Y})=\frac{1}{1+\mbox{AO}(\bx;P_{\,Y})} .
\end{equation*} 
To compute the finite-sample SPD we have to rely on approximate 
algorithms, as it is infeasible to consider all directions $\bv$. 
A convenient affine invariant procedure is obtained by 
considering directions $\bv$ which are orthogonal to an affine 
hyperplane through $p$ randomly drawn data points. 
In our implementation we use $250 p$ directions.
Table \ref{table:compAO} shows the time needed to compute 
the AO (or SPD) of $m \in \left\{1, 50, 100, 1000\right\}$ 
points with respect to a sample of $n=100$ points 
in dimensions $p \in \left\{2, 3, 4, 5\right\}$,
as in Table \ref{table:compbd}.
Here the time for $m=1$ is the fixed cost of computing those
$250 p$ directions and projecting the original data on them.

\begin{table}[htbp]
\centering
\caption{Computation times for the AO ($n=100$), 
         in units of 0.001 seconds.}
\label{table:compAO}
\vspace{0.2cm}
\begin{tabular}{|r|rrrr|}
\hline
  & \multicolumn{4}{c|}{$m$} \\
  $p$ & 1     & 50   & 100   & 1000  \\ 
\hline
2 & 15.0 & 15.3 & 15.6 & 20.9 \\
3 & 23.2 & 23.9 & 23.5 & 31.3 \\
4 & 30.5 & 30.9 & 31.6 & 41.7 \\
5 & 38.4 & 39.1 & 40.0 & 52.2 \\
\hline           
\end{tabular}
\end{table}

We see that computing AO is much faster than computing the bagdistance (Table~\ref{table:compbd}), and that this difference becomes more pronounced at larger $p$ and $m$. 
This is mainly due to the fact that AO does not require to 
compute the deepest point in multivariate space, unlike the 
bagdistance~\eqref{eq:bagdistanceDef} which requires $\btheta$.
 

\section{Multivariate classifiers}
\label{sec:MultivariateClassifiers} 
\subsection{Existing methods}
\label{sec:ExistingMethods}

One of the oldest nonparametric classifiers is the $k$-nearest 
neighbor (kNN) method introduced by \cite{Fix:Knn}.
For each new observation the method looks up the $k$ training 
data points closest to it (typically in Euclidean distance), 
and then assigns it to the most prevalent group among those 
neighbors.
The value of $k$ is typically chosen by cross-validation to
minimize the misclassification rate.

\cite{Liu:Simplicial} proposed to assign a new observation to
the group in which it has the highest depth.
This {\it MaxDepth} rule is simple and can be applied to more 
than two groups.
On the other hand it often yields ties when the depth function
is identically zero on large domains, as is the case with
halfspace depth and simplicial depth. 
\cite{Dutta:depth} avoided this problem by using projection 
depth instead, whereas \cite{Hubert:ClassifSkewed} employed the 
skew-adjusted projection depth.

To improve on the {\it MaxDepth} rule, \cite{Li:DDplot}
introduced the {\it DepthDepth} classifier as follows.
Assume that there are two groups, and denote the empirical
distributions of the training groups as $P_1$ and $P_2$.
Then transform any data point $\bx \in \mathbb{R}^p$ to the
bivariate point
\begin{equation}
\label{eq:depthTransform}
   (\mbox{\it depth}(\bx; P_1),\mbox{\it depth}(\bx; P_2)) 
\end{equation}
where {\it depth} is a statistical depth function. 
These bivariate points form the so-called 
{\it depth-depth plot}, in which the two groups of training 
points are colored differently.
The classification is then performed on this plot.
The {\it MaxDepth} rule corresponds to separating according 
to the 45 degree line through the origin, but in general 
\cite{Li:DDplot} calculate the best separating polynomial.
Next, they assign a new observation to group 1 if it
lands above the polynomial, and to group 2 otherwise.
Some disadvantages of the depth-depth rule are the computational
complexity of finding the best separating polynomial and the need 
for majority voting when there are more than two groups. 
Other authors carry out a depth transform followed by
linear classification \citep{Lange:Classification} or 
kNN \citep{Cuesta:DDG} instead.

\subsection{Classification in distance space}
\label{sec:ClassificationInDistanceSpace}
It has been our experience that distances can be very useful
in classification, but we prefer not to give up the affine
invariance that depth enjoys.
Therefore, we propose to use the bagdistance of 
Section \ref{sec:Bagdistance} for this purpose, or alternatively 
the adjusted outlyingness of 
Section \ref{sec:SkewAdjustedProjectionDepth}.
Both are affine invariant, robust against outliers in the training
data, and suitable also for skewed data.

Suppose that $G$ groups (classes) are given, where 
$G \geqslant 2$. 
Let $P_g$ represent the empirical distribution of the training 
data from group $g=1,\ldots,G$. 
Instead of the depth transform \eqref{eq:depthTransform} we
now carry out a {\it distance transform} by mapping each 
point $\bx \in \mathbb{R}^p$ to the $G$-variate point
\begin{equation}
\label{eq:distTransform}
  (\mbox{\it dist}(\bx; P_1),\ldots,\mbox{\it dist}(\bx; P_G)) 
\end{equation}
where $\mbox{\it dist}(\bx; P_g)$ is a generalized distance or
an outlyingness measure of the point $\bx$ to the $g$-th 
training sample.
Note that the dimension $G$ may be lower, equal, or higher
than the original dimension $p$.  
After the distance transform any 
multivariate classifier may be applied, such as linear or
quadratic discriminant analysis. 
The simplest version is of course {\it MinDist}, which just 
assigns $\bx$ to the group with smallest coordinate 
in \eqref{eq:distTransform}. 
When using the Stahel-Donoho or the adjusted outlyingness, 
this is equivalent to the {\it MaxDepth} rule based on 
projection depth or skew-adjusted projection depth. 
However, we prefer to apply kNN to the transformed
points.
This combines the simplicity and robustness of kNN with the
affine invariance offered by the transformation.
Also note that we never need to resort to majority voting.
In the simulations in Section \ref{sec:Simulation} we will see 
that the proposed {\it DistSpace} method (i.e.\ the distance
transform \eqref{eq:distTransform} followed by kNN) works 
quite well.

\begin{figure}[!ht]
\centering
\includegraphics[width=1\textwidth]
 	              {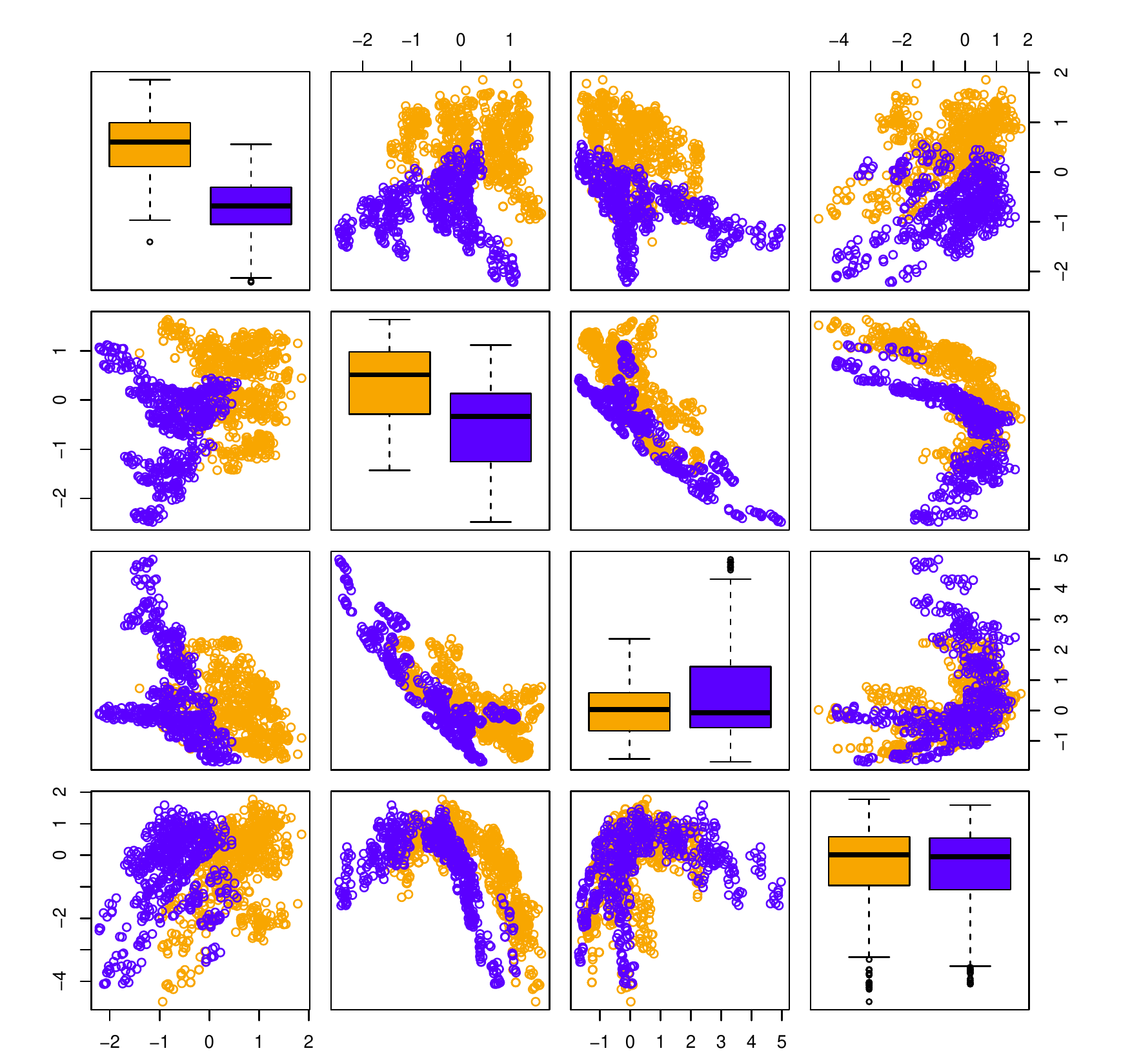}
\caption{Scatterplot matrix of the banknote authentication 
         data. The authentic bank-notes are shown in orange,
				 the forged ones in blue.}
\label{fig:Banknote_ScatterMatrix}
\end{figure}

We now illustrate the distance transform on a real world 
example, available from the UCI Machine Learning Repository 
\citep{Bache:UCIRepository}.
The data originated from an authentication procedure of banknotes. 
Photographs of 762 genuine and 610 forged banknotes were 
processed using wavelet transformations, and four features were 
extracted.
These are the 4 coordinates shown in the scatterplot matrix in 
Figure \ref{fig:Banknote_ScatterMatrix}. 

\begin{figure}[!ht]
\centering
\includegraphics[width=0.5\textwidth]{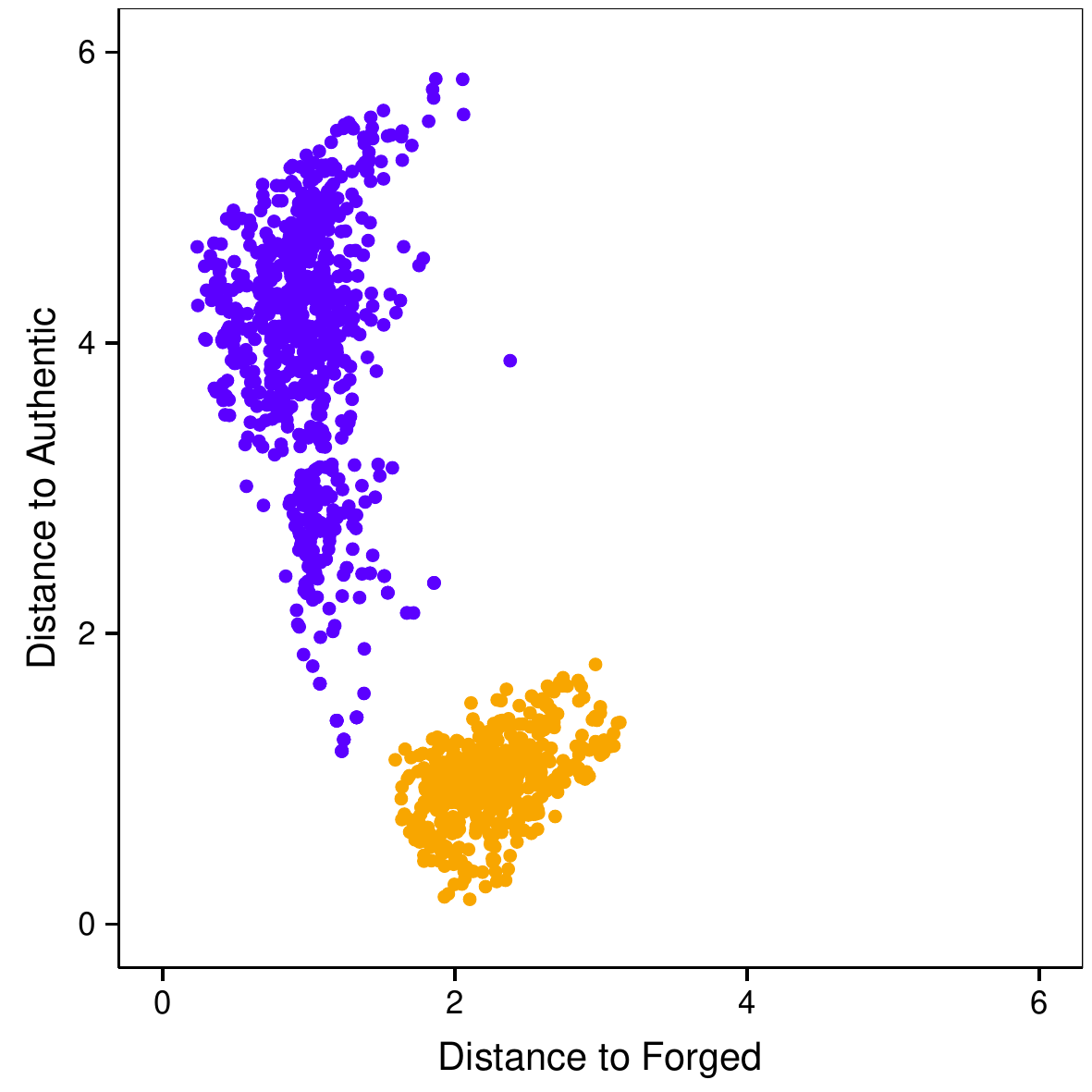}
\vspace{-0.2cm}
\caption{Distance-distance plot of the banknote 
         authentication data.}
\label{fig:Banknote_DistanceDistance}
\end{figure}

Note that $G=2$. Using the bagdistance, the distance space of 
this data is Figure~\ref{fig:Banknote_DistanceDistance}. 
It shows that forged and authentic banknotes are 
well-separated and that the authentic banknotes form a tight
cluster compared to that of the forged ones.
Any new banknote would yield a new point in this plot,
allowing kNN to classify it.


\section{Computational results} 
\label{sec:Simulation} 
To evaluate the various classifiers we apply them to simulated and real data.
Their performance is measured by their average 
misclassification percentage $\sum_{g=1}^{G} e_g n_g / N$ 
with $e_g$ the percentage of misclassified observations of 
group $g$ in the test set, $n_g$ the number of observations of
group $g$ in the training set, and $N$ the total size of the 
training set. This weights the misclassification percentages in the test set 
according to the prior probabilities. 
In each scenario the test set consists of 500 observations
per group.
This procedure is repeated 2000 times for each setting.

\textbf{Setting 1: Trivariate normals ($G=3,\;p=3$).} 
We generate data from three different normal distributions. 
The first group $C_1$ has parameters 
\[\bmu_1= \begin{pmatrix}   
					0 \\   0 \\   0  
					\end{pmatrix} 
					\qquad \text{and} \qquad
	\bSigma_1 = \begin{pmatrix}   
							5 & 3 & 1 \\
							3 & 2 & 1 \\
							1 & 1 & 3  
							\end{pmatrix} \;\;.
							\]
The second group is generated like $C_1$ but we flip the sign of 
the second coordinate. The third group is again generated like 
$C_1$ but then shifted by the vector $\left(1,-2,-4\right)$. 
The training data consist of 50 observations in each group.

\textbf{Setting 2: Multivariate normal and skewed ($G=2,\; p=6$).}
We consider two 6-variate distributions. 
The first group $C_1$ is drawn from the standard normal 
distribution. 
The coordinates in the second group are independent draws 
from the exponential distribution with rate parameter 1: 
\[C_1 \sim \mbox{N}\left(\bzero, \boldsymbol{I}_6 \right) \quad \text{ and }
 \quad C_2 \sim  (\mbox{Exp}(1),\mbox{Exp}(1),\mbox{Exp}(1),
 \mbox{Exp}(1),\mbox{Exp}(1),\mbox{Exp}(1))'. \]
The training data has 150 observations drawn from group $C_1$ 
and 100 from $C_2$.

\textbf{Setting 3: Concentric distributions ($G=2, p=7$).}
This consists of two groups of data. 
The first group is drawn from the standard normal
distribution. 
The second group is obtained by generating points on the
unit sphere in $\mathbb{R}^p$ and multiplying them by
lengths which are generated uniformly on $[12,13]$.
The training data has 150 observations from group $C_1$ 
and 250 from $C_2$.

\textbf{Setting 4: Banknote authentication data ($G=2,\;p=4$).}
We first standardize the data by the columnwise 
median and $\mbox{MAD}$. 
The training sets are random subsets of 500 points from the 
original data set, with the test sets each time consisting of 
the remaining 872 observations. 

Among the depth-based classification rules, halfspace depth (HD) 
is compared to projection depth (PD) and skew-adjusted 
projection depth (SPD). We run the {\it MaxDepth} rule,  
{\it DepthDepth} followed by the best separating polynomial and
{\it DepthDepth} followed by kNN. The degree of the polynomial and the number of neighbors $k$ are selected based on leave-one-out cross-validation. 
 
Among the distance-based classifiers, the bagdistance based on halfspace depth~({\it bd\,}) 
is compared to the Stahel-Donoho outlyingness (SDO) and the 
adjusted outlyingness (AO). Here the {\it MinDist} and {\it DistSpace} 
classifiers are considered. 

\begin{figure}[!ht]
\centering
\includegraphics[width=1\textwidth]{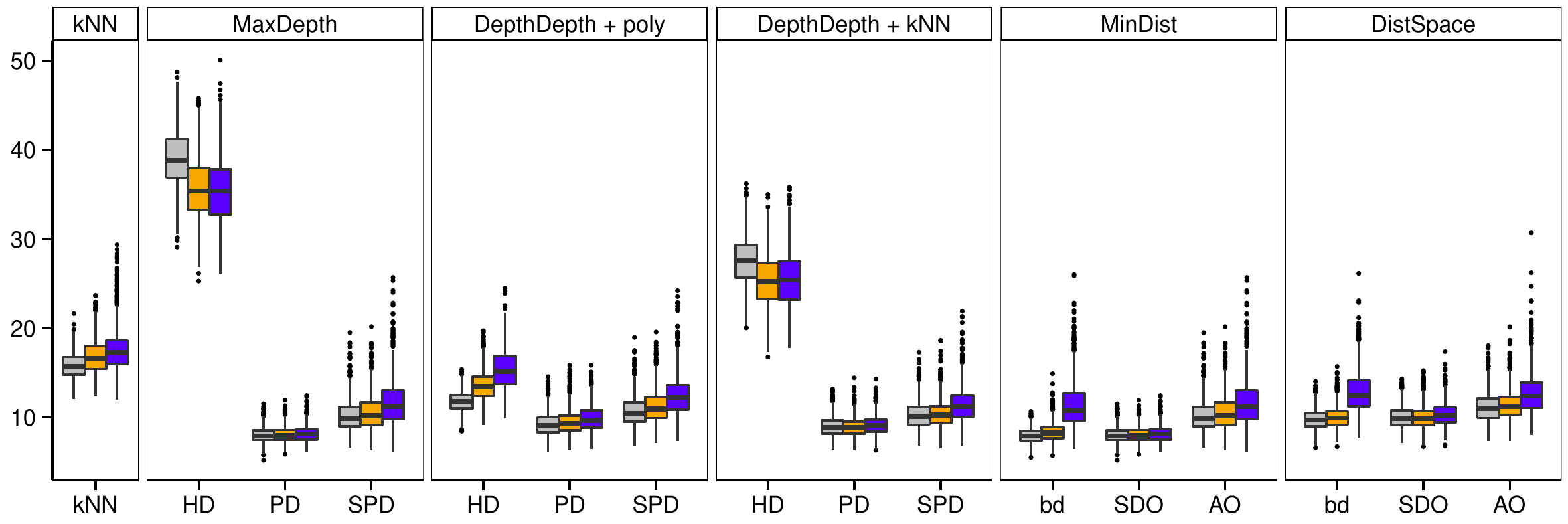}
\caption{Misclassification percentages in 2000 runs of setting 1 
         (trivariate normals). The results for clean data 
				 are shown in gray, for 5\% mislabeled data in orange,
				 and for 10\% mislabeled data in blue.}
\label{fig:MScenario_Norm}
\end{figure}

\begin{figure}[!ht] 
\centering
\includegraphics[width=1\textwidth]{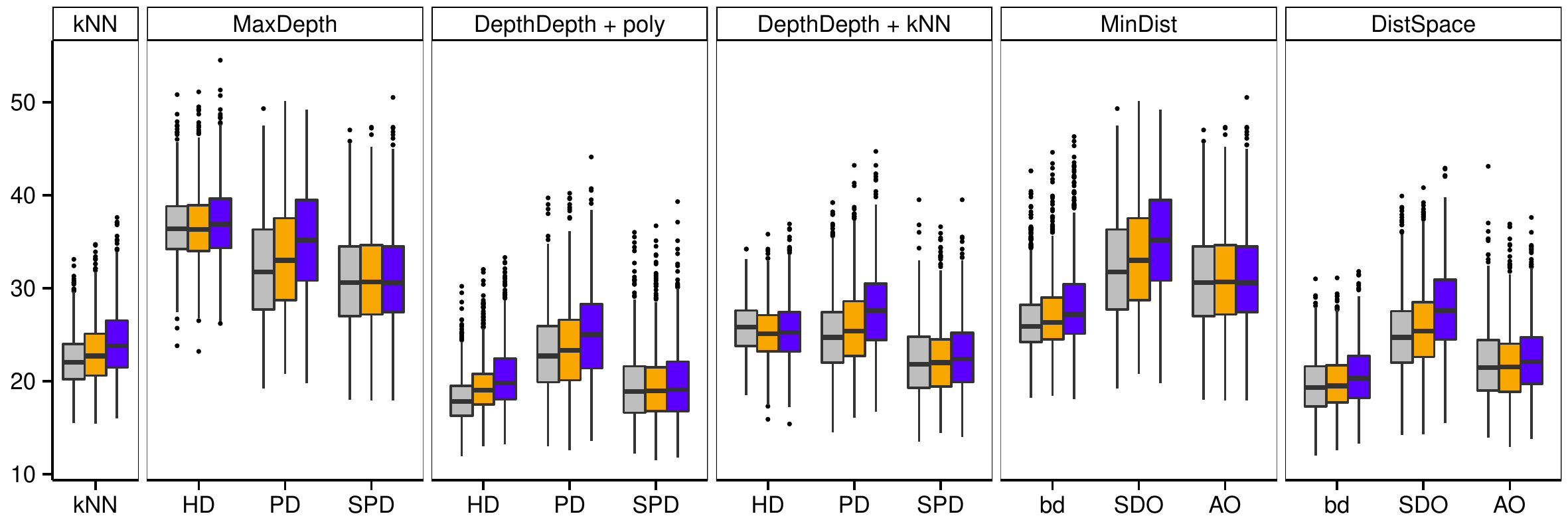}
\caption{Misclassification percentages in 2000 runs of setting 2
         (6-variate normal and skewed) using the same color code.}
\label{fig:MScenario_NormExp}
\end{figure}

\begin{figure}[!ht] 
\centering
\includegraphics[width=1\textwidth]{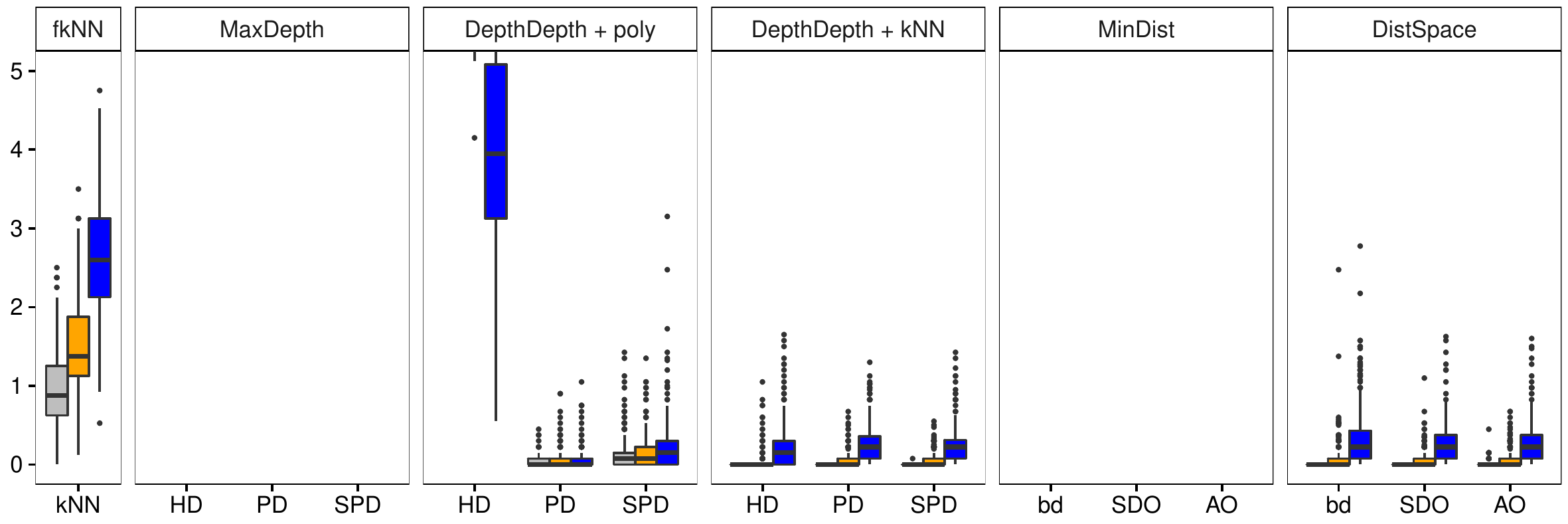}
\caption{Misclassification percentages in 2000 runs of 
         setting 3 (concentric groups).}
\label{fig:MScenario_Pencil}
\end{figure}

\begin{figure}[!ht] 
\centering
\includegraphics[width=1\textwidth]{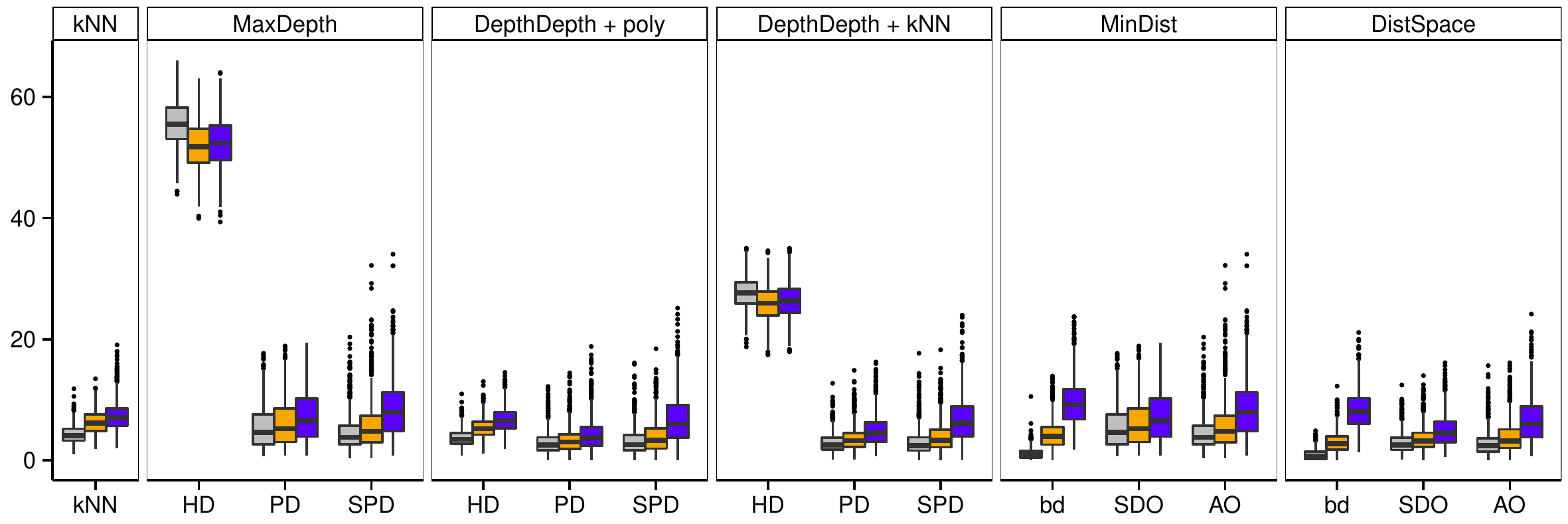}
\caption{Misclassification percentages in 2000 runs of setting 4
         (banknote data).}
\label{fig:MScenario_Bank}
\end{figure}

We evaluate all classifiers on the uncontaminated data, and on data 
where 5\% and 10\% of the observations in each group are mislabeled by 
assigning them randomly to another group. 
Figures~\ref{fig:MScenario_Norm}-\ref{fig:MScenario_Bank}
summarize the results with boxplots of the misclassification percentages.
 
In setting 1, most of the depth- and distance-based methods
did better than kNN. The halfspace depth HD did not perform 
well in {\it MaxDepth} and {\it DepthDepth + kNN}, and in fact 
mislabeling improved the classification because it yielded 
fewer points with depth zero in both groups. 
Halfspace depth appeared to work better in {\it DepthDepth + 
polynomial} but this is due to the fact that whenever a
point has depth zero in both groups, \cite{Li:DDplot}
fall back on kNN in the original data space. Also note
that {\it DepthDepth + polynomial} by construction 
improves the {\it MaxDepth} rule on training data, but it doesn't
always perform better on test data. 

In setting 2 we note the same things about HD in the depth-based methods. 
The best results are obtained by {\it DepthDepth + poly}
and {\it DistSpace}, where we note that the methods that are 
able to reflect skewness (HD, SPD, {\it bd}, AO) did a lot better
than those that aren't (PD, SDO).
This is because the data contains a skewed group.
\vspace{0.2cm}

In the third setting one of the groups is not convex at all,
and the {\it MaxDepth} and {\it MinDist} boxplots lie
entirely above the figure. 
On the other hand the {\it DepthDepth} and {\it DistSpace}
methods still see structure in the data, and yield better 
results than kNN on the original data.

In the banknote authentication example (setting 4), all
methods except HD work well. For clean data, the two methods
using the bagdistance outperform all others.


\section{Functional data}
\label{sec:FunctionalDataAndItsTools}

The analysis of functional data is a booming research area of 
statistics, see e.g.\ the books of \cite{Ramsay:BookFDA} and 
\cite{Ferraty:BookFDA}. 
A functional data set typically consists of $n$ curves observed 
at time points $t_1, \ldots, t_T$. 
The value of a curve at a given time point is a $p$-variate
vector of measurements.
We call the functional dataset univariate or multivariate 
depending on $p$.
For instance, the multi-lead ECG data set analyzed by 
\cite{Pigoli:Wavelets} is multivariate with $p=8$.

When faced with classification of functional data, one approach
is to consider it as multivariate data in which the 
measurement(s) at different time points are separate variables. 
This yields high-dimensional data with typically many highly 
correlated variables, which can be dealt with by
penalization \citep{hastie1995}. 
Another approach is to project such data onto a lower-dimensional 
subspace and to continue with the projected data,
e.g. by means of support vector 
machines \citep{Rossi:SVM_FDA, MartinBarragan:InterSVM_FDA}. 
\cite{Li:SegmentationApproach} proposed to use 
$F$-statistics to select small subintervals in the domain and 
to restrict the analysis to those.
Other techniques include the weighted distance method of
\cite{Alonso:WeightedDistance} and the componentwise approach
of \cite{Delaigle:ClassifFunct}. 

To reflect the dynamic behavior of functional data one can add
their derivatives or integrals to the analysis, and/or add
some preprocessing functions (warping functions, baseline 
corrections, \ldots) as illustrated in \cite{Claeskens:MFHD}. 
This augments the data dimension and may add valuable 
information that can be beneficial in obtaining a better
classification. 
We will illustrate this on a real data set in 
Section~\ref{sec:Writing}.
     
The study of robust methods for functional data started only 
recently. 
So far, efforts to construct robust classification rules for 
functional data have mainly used the concept of depth:
\cite{Lopez:ClassifDepth} used the modified band depth,
\cite{Cuesta:FCRTD} made use of random Tukey depth, and
\cite{Hlubinka:ClassifDepth} compared several depth 
functions in this context.

\subsection{Functional depths and distances}
\label{sec:FunctionalDepthsAndDistances}
\cite{Claeskens:MFHD} proposed a type of
multivariate functional depth (MFD) as follows.
Consider a $p$-variate stochastic process  
$Y=\left\{Y(t), t \in U \right\}$, a statistical depth function
$D(\cdot,\cdot)$ on $\mathbb{R}^p$, and a weight function $w$ 
on $U$ integrating to 1. 
Then the MFD of a curve $X$ on $U$ with respect to the 
distribution $P_{\,\,Y}$ is defined as 
\begin{equation}
  \label{eq:mfd}
  \mbox{MFD}(X;P_{\,Y}) =
	\int_{U}{D(X(t); P_{\,Y(t )}) \, w(t) \, dt} 
\end{equation} 
where $P_{\,Y(t)}$ is the distribution of $Y$ at time $t$. 
The weight function $w(t)$ allows to emphasize or downweight 
certain time regions, but in this paper will be assumed
constant. 
The functional median $\Theta(t)$ is defined as the curve 
with maximal MFD. 
Properties of the MFD may be found in \citep{Claeskens:MFHD}, 
with emphasis on the case where $D(\cdot,\cdot)$ is the 
halfspace depth. Several consistency results are derived in
\citep{Nagy:ConsistencyID}.
 
For ease of notation and to draw quick parallels to the 
multivariate non-functional case, we will denote the MFD based 
on halfspace depth by fHD, and the MFD based on projection 
depth and skew-adjusted projection depth by fPD and fSPD.
 
Analogously, we can define the {\it functional bagdistance} 
{\it (fbd)} of a curve $X$ to (the distribution of) a 
stochastic process $Y$ as 
\begin{equation} 
\label{eq:fbd} 
  \mbox{\it fbd}(X;P_{\,Y}) =
	\int_{U}{bd(X(t); P_{\,Y}(t)) \, dt}\; .  
\end{equation}
Similar extensions of the Stahel-Donoho outlyingness SDO 
and the adjusted outlyingness AO to the functional context 
are given by
\begin{eqnarray} 
  \label{eq:fSDO}
  \mbox{fSDO}(X; P_{\,Y}) =
	   \int_U {\mbox{SDO} (X(t); P_{\,Y}(t)) \; dt} \; \\
	\label{eq:fAO}
  \mbox{fAO}(X; P_{\,Y})  =
	   \int_U {\mbox{AO}(X(t); P_{\,Y}(t)) \; dt} \; .
\end{eqnarray}

\subsection{Functional classifiers} 
\label{sec:FunctionalClassifiers}

The classifiers discussed in 
Section \ref{sec:MultivariateClassifiers} are readily adapted 
to functional data. 
By simply plugging in the functional versions of the distances 
and depths all procedures can be carried over. 
For the $k$-nearest neighbor method one typically uses the
$L^2$-distance: 
$$d_2(X_1,X_2) = \left( \int_{U}{\norm{X_1(t)-X_2(t)}^2 \, dt}
  \right)^{1/2}\;\; .$$
The functional kNN method will be denoted as fkNN. 
It is simple but not affine invariant.
Analogously we use the {\it MaxDepth} and {\it DepthDepth} 
rules based on fHD, fPD, and fSPD, as well as the 
{\it MinDist} and {\it DistSpace} rules based on {\it fbd}, 
fSDO, and fAO. 
Note that \cite{Mosler:DDClass} already studied {\it DepthDepth} on
functional data after applying a dimension reduction 
technique. 


\section{Functional data examples}
\label{sec:FunctionalExamples}

\subsection{Fighter plane dataset} 
\label{sec:Plane} 

\begin{figure}[!ht]
\centering 		
\includegraphics[width=1\textwidth]{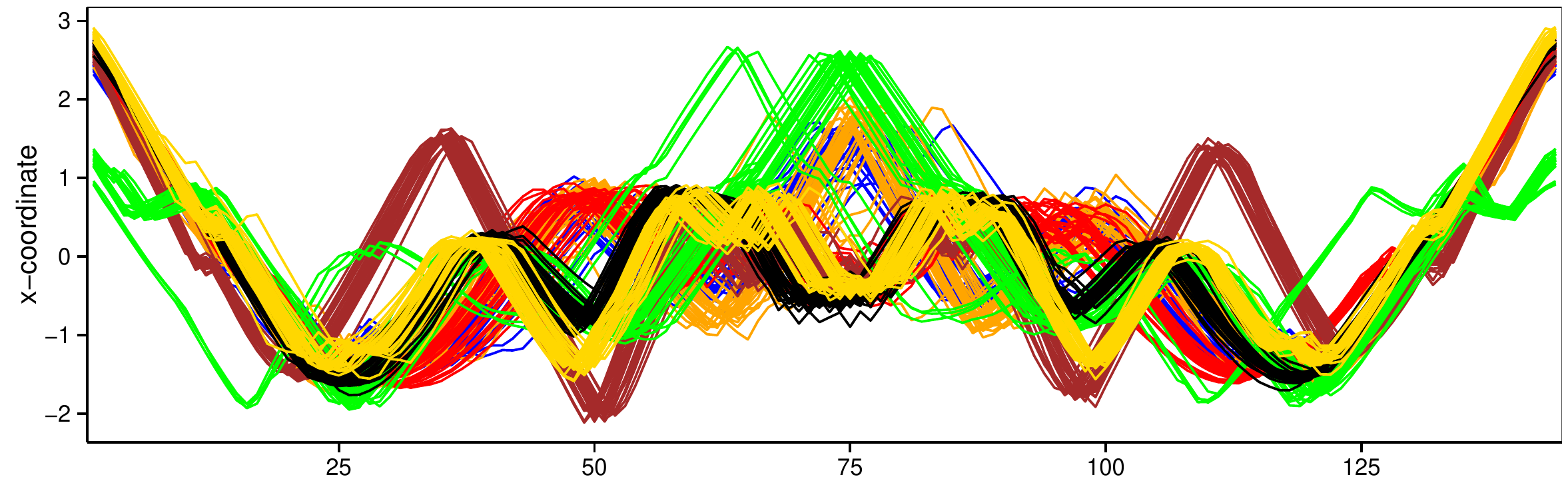}
\caption{Functions describing the shapes of fighter 
         planes.}
\label{fig:Plane_curves}
\end{figure}

The fighter plane dataset of~\cite{Thakoor:Planes}
describes 7 shapes: of the Mirage, Eurofighter,
F-14 with wings closed, F-14 with wings opened, Harrier,
F-22 and F-15.
Each class contains 30 shape samples obtained from
digital pictures, which~\cite{Thakoor:Planes} then 
reduced to the univariate functions in 
Figure~\ref{fig:Plane_curves}.
We obtained the data from the UCR Time Series Classification 
Archive~\citep{UCR:Archive}. 

\begin{figure}[!ht]
\centering 		
\includegraphics[width=1\textwidth]{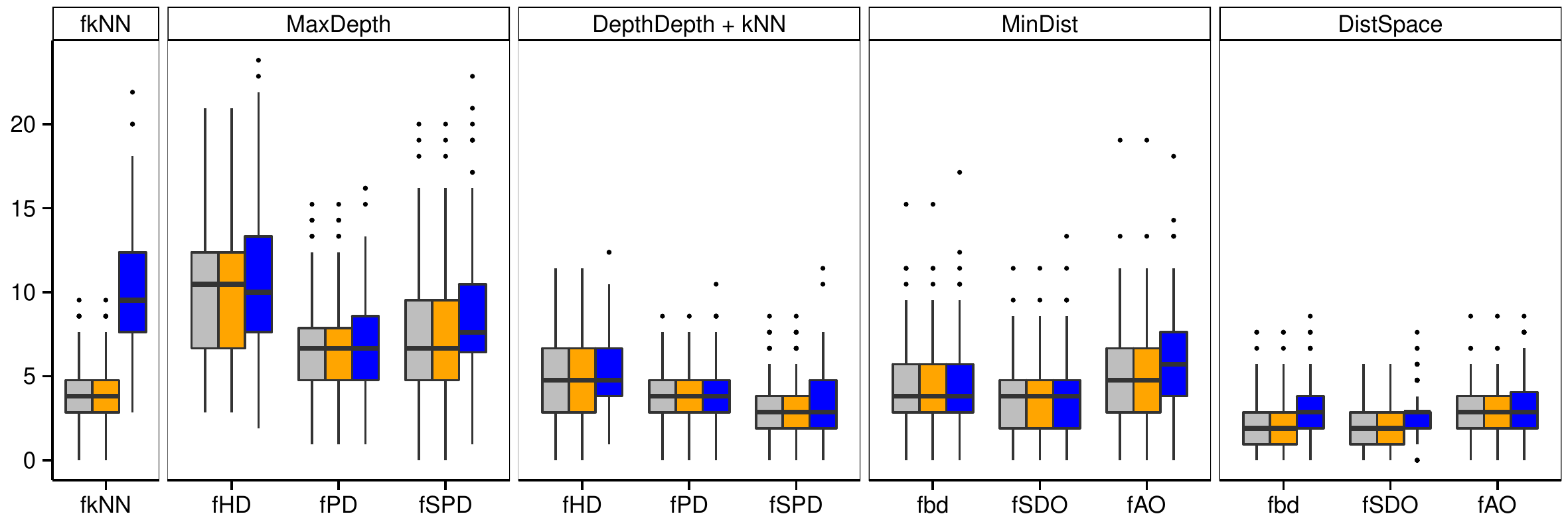}
\caption{Misclassification percentages in 200 runs of the 
         fighter plane data.}
\label{fig:Plane_boxplots}
\end{figure}

In all, the plane data set consists of 210 observations 
divided among 7 groups. 
For the training data we randomly drew 15 observations from 
each group, and the test data were the remaining 105 
observations.
Repeating this 200 times yielded the misclassification 
percentages in Figure~\ref{fig:Plane_boxplots}. 

In this data set the {\it DistSpace} method performed
best, followed by kNN which however suffered under 10\%
of mislabeling.
Figure~\ref{fig:Plane_boxplots} contains no panel for
{\it DepthDepth + poly} because the computation time of
this method was infeasible due to the computation of
the separating polynomials combined with majority voting 
for $G=7$. 

\subsection{MRI dataset} 
\label{sec:MRI} 
\begin{figure}[!ht]
\centering 		
\includegraphics[width=1\textwidth]{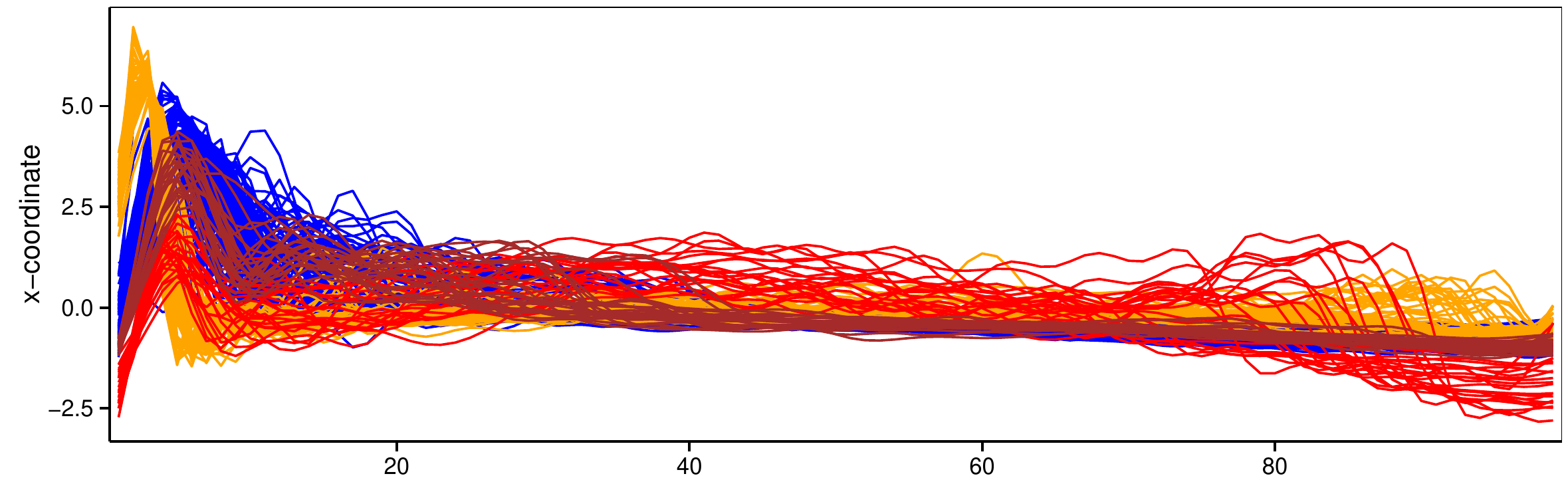} 
\caption{Curves computed from MRI intensities.}
\label{fig:MRI_curves}
\end{figure}

\cite{Felipe:MRI} obtained intensities of MRI images of
9 different parts of the human body (plus a group 
consisting of all remaining body regions, which was of
course very heterogeneous).
They then transformed their data to curves.
This data set was also downloaded from~\citep{UCR:Archive}. 
The $G=9$ classes together contain 547 observations and are of
unequal size. For example $n_1=112, n_2=65, n_3=75,...$.
The curves for 4 of these classes are shown in 
Figure~\ref{fig:MRI_curves} (if we plot all 9 groups
together, some become invisible).

\begin{figure}[!ht]	
\centering 		
\includegraphics[width=1\textwidth]{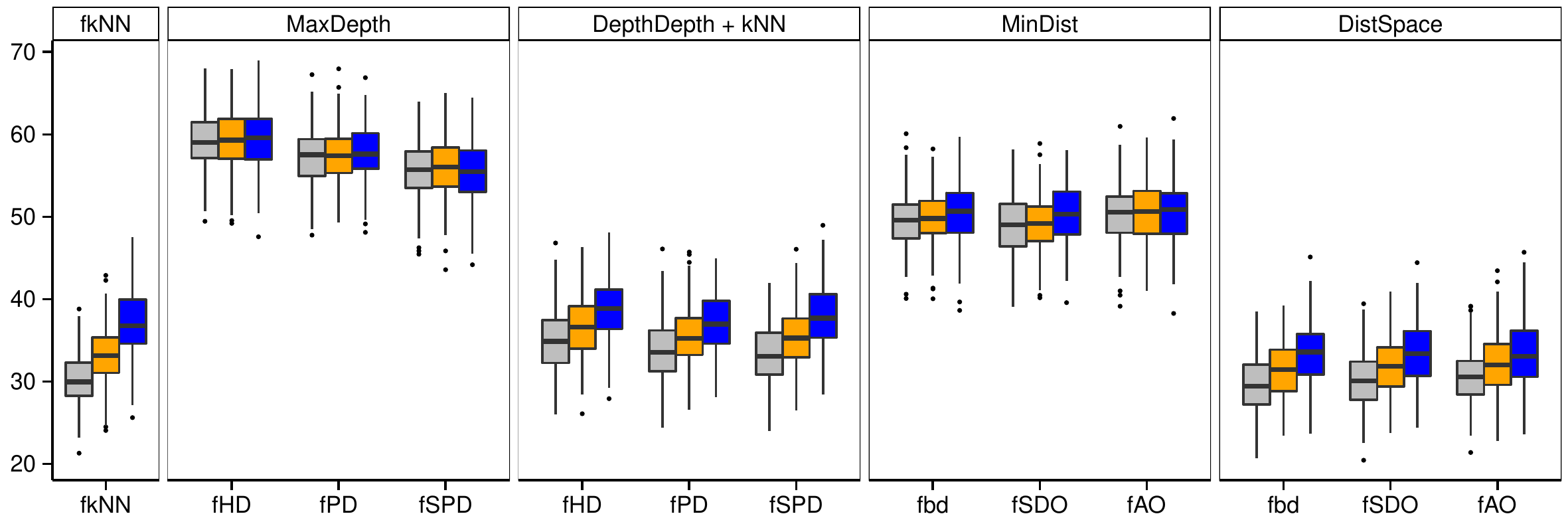}
\caption{Misclassification percentages in 200 runs of 
         the MRI data.}
\label{fig:MRI_boxplots}
\end{figure}

For the training data we drew unequally sized random subsets 
from these groups. 
The misclassification rates of 200 experiments of this
type are shown in Figure~\ref{fig:MRI_boxplots}. 


Here {\it DistSpace} performs a bit better than fKNN
under contamination, and much better than {\it MaxDepth}
and {\it MinDist}. Also in this example the 
{\it DepthDepth + poly} method took too long to compute.

\subsection{Writing dataset}
\label{sec:Writing}

\begin{figure}[!ht]
\centering
\includegraphics[width=1\textwidth]{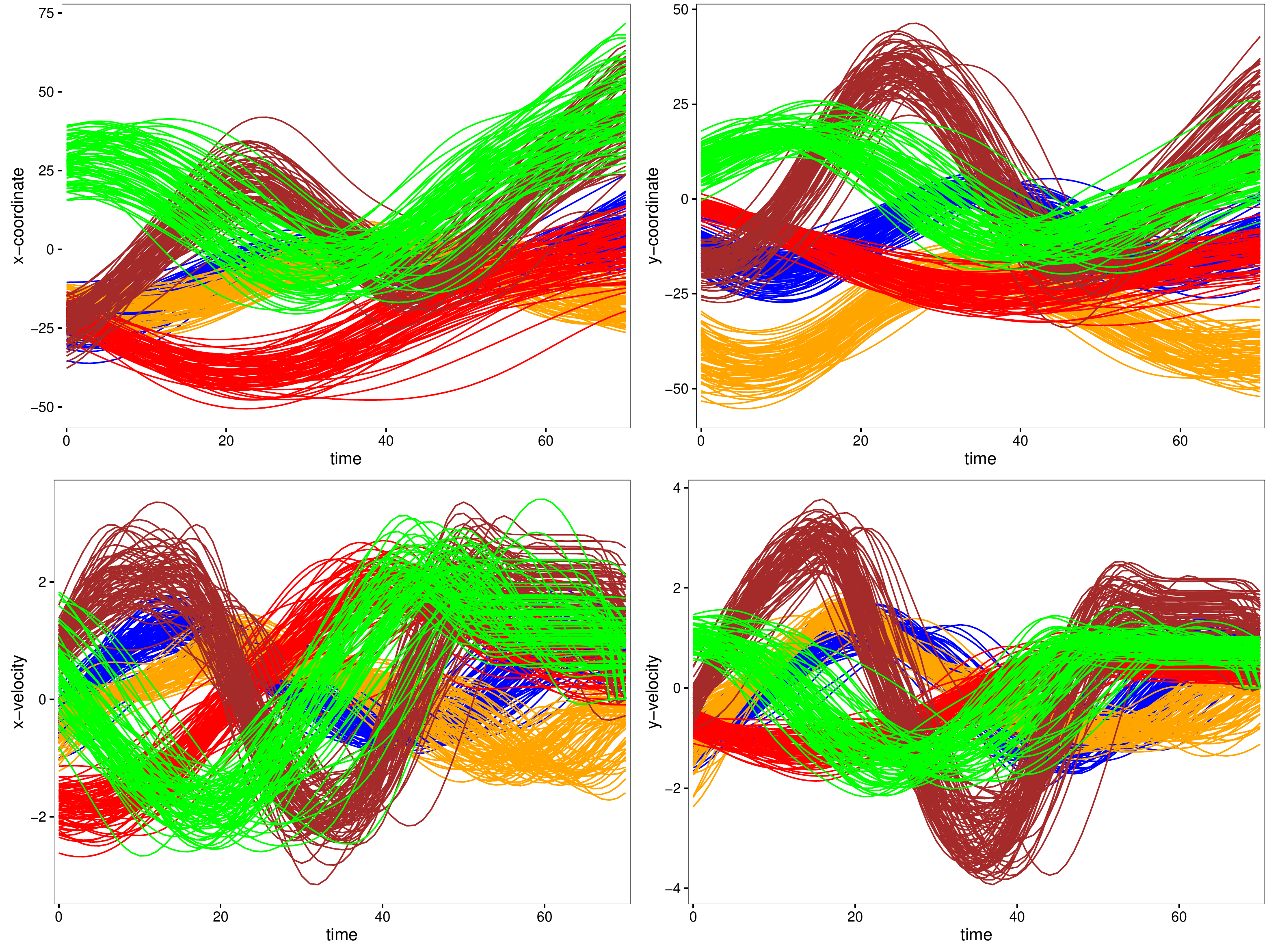}
\caption{Coordinates (upper) and speed (lower) of the 
         writing data.
         Each group has a different color.}
\label{fig:WritingPlot}
\end{figure}

The writing dataset consists of 2858 character samples 
corresponding to the speed profile of the tip of a pen writing 
different letters, as captured on a WACOM tablet. 
The data came from the UCI Machine Learning Repository  
\citep{Bache:UCIRepository}.
We added the $x$- and $y$-coordinates of the pen tip 
(obtained by integration) to the data, yielding $p=4$ overall
unlike both previous examples which had $p=1$. 
We further processed the data by removing the first and last
time points and by interpolating to give all curves the same 
time domain. 
Samples corresponding to the letters `a', `c', `e', `h' and `m' 
were retained. 
This yields a five-group supervised classification problem of 
four-dimensional functional data. 
Figure \ref{fig:WritingPlot} plots the 
curves, with the 5 groups shown in different colors. 

\begin{figure}
\centering 
\includegraphics[width=1\textwidth]{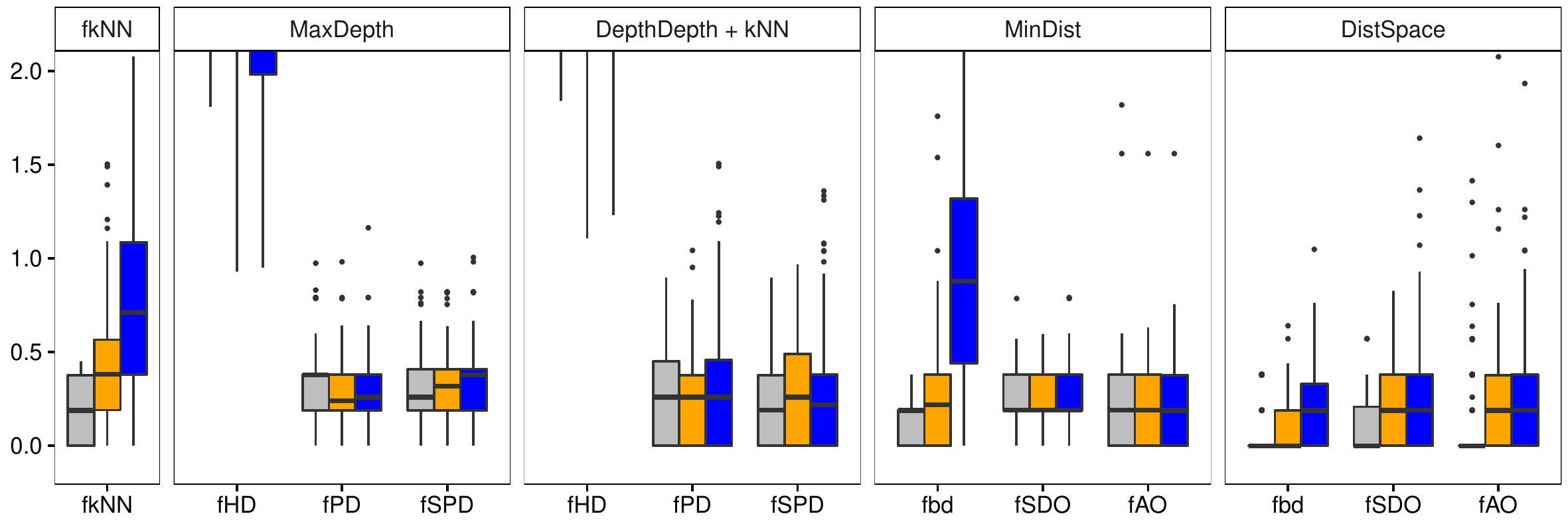}
\caption{Misclassification percentages in 200 runs of the 
         writing data.}
\label{fig:Letters}
\end{figure}

For each letter the training set was a random subset of 80 
multivariate curves. 
The outcome is in Figure~\ref{fig:Letters}. 
There is no panel for the {\it DepthDepth + poly} classifier 
with separating polynomials and majority voting as its
computation time was infeasible.
{\it MaxDepth} and {\it DepthDepth} combined with kNN 
perform well except for fHD, again due to the fact that HD
is zero outside the convex hull.
{\it DistSpace} outperforms {\it MinDist}, and works well
with all three distances. 
The best result was obtained by {\it DistSpace} with {\it fbd}.

Finally we applied fkNN and {\it DistSpace} to the original 
two-dimensional velocity data only. 
This resulted in larger median misclassification errors for 
all methods and all 3 data settings (0\%, 5\% and 10\% 
mislabeling). 
For example, {\it DistSpace} with {\it fbd} on the 
two-dimensional data yielded a median misclassification error 
of 0.35\%, whereas the median error was zero on the
4-dimensional augmented data. 
This shows that adding appropriate data-based functional 
information can be very useful to better separate groups.   
%
%


\section{Conclusions}
\label{sec:Conclusion}

Existing classification rules for multivariate or functional
data, like kNN, often work well but can fail when the  
dispersion of the data depends strongly on the direction
in which it is measured.
The {\it MaxDepth} rule of \cite{Liu:Simplicial} and its
{\it DepthDepth} extension \citep{Li:DDplot} resolve this
by their affine invariance, but perform poorly in
combination with depth functions that become zero outside 
the convex hull of the data, like halfspace depth (HD).

This is why we prefer to use the bagdistance {\it bd}, 
which is based on HD and has properties very close to those 
of a norm but is able to reflect skewness (while still
assuming some convexity). 
Rather than transforming the data to their depths we propose
the {\it distance transform}, based on {\it bd} or a 
measure of outlyingness such as SDO or AO.

After applying the depth or distance transforms there
are many possible ways to classify the transformed data.
We found that the original separating polynomial method did 
not perform the best.
Therefore we prefer to apply kNN to the transformed data.

In our experiments with real and simulated data we found
that the best performing methods overall were 
{\it DepthDepth + kNN} (except with halfspace depth)
and {\it DistSpace + kNN}.
The latter approach combines affine invariance with the
computation of a distance and the simplicity, lack of 
assumptions, and robustness of kNN,
and works well for both multivariate and functional data.

In the multivariate classification setting the depth
and distance transforms perform about equally well, and
in particular {\it MinDist} on SDO and AO is equivalent
to {\it MaxDepth} on the corresponding depths PD and SPD.
But the bagdistance {\it bd} beats the
halfspace depth HD in this respect because the latter
is zero outside the convex hull of a group.

One of the most interesting results of our simulations is
that the depth and distance transforms are less similar
in the functional setting. 
Indeed, throughout Section \ref{sec:FunctionalExamples}
the distance transform outperformed the depth transform.
This is because distances are more additive than depths,
which matters because of the integrals in the definitions
of functional depth MFD~\eqref{eq:mfd} to functional
AO~\eqref{eq:fAO}.
For the sake of simplicity, let us focus on the empirical 
versions where the integrals in~\eqref{eq:mfd} 
to~\eqref{eq:fAO} become sums over a finite number of
observed time points.
These sums are $L^1$ norms (we could also use $L^2$ norms 
by taking the square root of the sum of squares).
In the context of classification, we are measuring how
different a new curve $X$ is from a process $Y$ or
a finite sample from it.
When $X$ differs strongly from $Y$ in a few time
points, the integrated depth~\eqref{eq:mfd} will have a
few terms equal to zero or close to zero, which will
not lead to an extremely small sum, so $X$ would appear 
quite similar to $Y$.
On the other hand, a functional distance measure
like~\eqref{eq:fbd}--\eqref{eq:fAO}
will contain a few very large terms, which will have a
large effect on the sum, thereby revealing that $X$ is
quite far from $Y$.
In other words, functional distance adds up information
about how distinct $X$ is from $Y$.
The main difference between the two approaches is that the 
depth terms are bounded from below (by zero), whereas 
distance terms are unbounded from above and thus better 
able to reflect discrepancies.

\bibliographystyle{apalike} 

\begin{thebibliography}{}

\bibitem[Alonso et~al., 2012]{Alonso:WeightedDistance}
Alonso, A., Casado, D., and Romo, J. (2012).
\newblock Supervised classification for functional data: a 
weighted distance approach.
\newblock {\em Computational Statistics \& Data Analysis}, 
56:2334--2346.

\bibitem[Bache and Lichman, 2013]{Bache:UCIRepository}
Bache, K. and Lichman, M. (2013).
\newblock {UCI} {M}achine {L}earning {R}epository, {\it https://archive.ics.uci.edu/ml/datasets.html}

\bibitem[Brys et~al., 2005]{Brys:RobICA}
Brys, G., Hubert, M., and Rousseeuw, P.~J. (2005).
\newblock A robustification of independent component analysis.
\newblock {\em Journal of Chemometrics}, 19:364--375.

\bibitem[Brys et~al., 2004]{Brys:RobSkew}
Brys, G., Hubert, M., and Struyf, A. (2004).
\newblock A robust measure of skewness.
\newblock {\em Journal of Computational and Graphical Statistics},
  13:996--1017.

\bibitem[Chen et~al., 2015]{UCR:Archive}
Chen, Y., Keogh, E., Hu, B., Begum, N., Bagnall, A., Mueen, A., Batista, G.J. (2015).
\newblock The UCR Time Series Classification Archive.
\newblock {\it www.cs.ucr.edu/$\sim$eamonn/time\_series\_data/}

\bibitem[Christmann et~al., 2002]{Christmann:RegrDeptSVM}
Christmann, A., Fischer, P., and Joachims, T. (2002).
\newblock Comparison between various regression depth methods
  and the support vector machine to approximate the minimum
	number of misclassifications.
\newblock {\em Computational Statistics}, 17:273--287.

\bibitem[Christmann and Rousseeuw, 2001]{Christmann:Overlap}
Christmann, A. and Rousseeuw, P.~J. (2001).
\newblock Measuring overlap in logistic regression.
\newblock {\em Computational Statistics \& Data Analysis}, 
          37:65--75.

\bibitem[Claeskens et~al., 2014]{Claeskens:MFHD}
Claeskens, G., Hubert, M., Slaets, L., and Vakili, K. (2014).
\newblock Multivariate functional halfspace depth.
\newblock {\em Journal of the American Statistical Association},
  109(505):411--423.

\bibitem[Cuesta-Albertos and Nieto-Reyes, 2010]{Cuesta:FCRTD}
Cuesta-Albertos, J.A. and Nieto-Reyes, A. (2010).
\newblock Functional classification and the random Tukey depth: Practical issues. 
\newblock In: Borgelt, C., Rodr\'iguez, G.G., Trutschnig, W.,
 Lubiano, M.A., Angeles Gil, M., Grzegorzewski, P. and 
 Hryniewicz, O. (eds.) {\em {C}ombining {S}oft {C}omputing and
 {S}tatistical {M}ethods in {D}ata {A}nalysis.} Springer, 
 Berlin Heidelberg, pages 123--130.

\bibitem[Cuesta-Albertos et~al., 2015]{Cuesta:DDG}
Cuesta-Albertos, J.A., Febrero-Bande, M., and Oviedo de la
Fuente, M. (2015).
\newblock The $DD^G$-classifier in the functional setting. 
\newblock arXiv:1501.00372v2.

\bibitem[Delaigle et~al., 2012]{Delaigle:ClassifFunct} 
Delaigle, A., Hall, P., and Bathia, N. (2012).
\newblock Componentwise classification and clustering of 
functional data.
\newblock {\em Biometrika}, 99:299--313.

\bibitem[Donoho, 1982]{Donoho:SDest}
Donoho, D. (1982).
\newblock {\em Breakdown properties of multivariate location estimators}.
\newblock Ph.D. Qualifying paper, Dept. Statistics, Harvard
  University, Boston.

\bibitem[Donoho and Gasko, 1992]{Donoho:Depth}
Donoho, D. and Gasko, M. (1992).
\newblock Breakdown properties of location estimates based on
 halfspace depth and projected outlyingness.
\newblock {\em The Annals of Statistics}, 20(4):1803--1827.

\bibitem[Dutta and Ghosh, 2011]{Dutta:depth}
Dutta, S. and Ghosh, A. (2011).
\newblock On robust classification using projection depth.
\newblock {\em Annals of the Institute of Statistical
   Mathematics}, 64:657--676.

\bibitem[Dyckerhoff and Mozharovskyi, 2016]{Dyckerhoff:HSdepth}
Dyckerhoff, R. and Mozharovskyi, P. (2016).
\newblock Exact computation of the halfspace depth. 
\newblock {\em Computational Statistics \& Data Analysis}, 98:19--30. 

\bibitem[Ferraty and Vieu, 2006]{Ferraty:BookFDA}
Ferraty, F. and Vieu, P. (2006).
\newblock {\em Nonparametric Functional Data Analysis: Theory
               and Practice}.
\newblock Springer, New York.

\bibitem[Felipe et~al., 2005]{Felipe:MRI}
Felipe, J.C., Traina, A.J.M., Traina, C. (2005).
\newblock Global warp metric distance: boosting content-based 
image retrieval through histograms.
\newblock Proceedings of the Seventh IEEE International 
          Symposium on Multimedia (ISM'05), p.8.

\bibitem[Fix and Hodges, 1951]{Fix:Knn}
Fix, E. and Hodges, J.~L. (1951).
\newblock Discriminatory analysis - nonparametric discrimination:
 Consistency properties.
\newblock {\em Technical Report 4 USAF School of Aviation 
 Medicine, Randolph Field, Texas}.

\bibitem[Ghosh and Chaudhuri, 2005]{Ghosh:Maxdepth}
Ghosh, A. and Chaudhuri, P. (2005).
\newblock On maximum depth and related classifiers.
\newblock {\em Scandinavian Journal of Statistics},
  32(2):327--350.

\bibitem[Hallin et~al., 2010]{Hallin:Quantiles}
Hallin, M., Paindaveine, D., and {\v{S}}iman, M. (2010).
\newblock Multivariate quantiles and multiple-output regression
 quantiles: from {$L\sb 1$} optimization to halfspace depth.
\newblock {\em The Annals of Statistics}, 38(2):635--669.

\bibitem[Hastie et~al., 1995]{hastie1995}
Hastie, T., Buja, A., and Tibshirani, R. (1995).
\newblock Penalized discriminant analysis.
\newblock {\em The Annals of Statistics}, 23(1):73--102.

\bibitem[Hastie et~al., 2009]{Hastie:StatLearning}
Hastie, T., Tibshirani, R., and Friedman, J. (2009).
\newblock {\em The Elements of Statistical Learning}.
\newblock Springer, New York, second edition.

\bibitem[Hlubinka et~al., 2015]{Hlubinka:ClassifDepth}
Hlubinka, D., Gijbels, I., Omelka, M., and Nagy, S. (2015).
\newblock Integrated data depth for smooth functions and
 its application in supervised classification.
\newblock {\em Computational Statistics}, 30:1011--1031.

\bibitem[Hubert et~al., 2015]{Hubert:MFOD}
Hubert, M., Rousseeuw, P.~J., and Segaert, P. (2015).
\newblock Multivariate functional outlier detection.
\newblock {\em Statistical Methods \& Applications}, 
           24:177--202.

\bibitem[Hubert and Van~der Veeken, 2010]{Hubert:ClassifSkewed}
Hubert, M. and Van~der Veeken, S. (2010).
\newblock Robust classification for skewed data.
\newblock {\em Advances in Data Analysis and Classification},
 4:239--254.

\bibitem[Hubert and Vandervieren, 2008]{Hubert:AdjBoxplot}
Hubert, M. and Vandervieren, E. (2008).
\newblock An adjusted boxplot for skewed distributions.
\newblock {\em Computational Statistics \& Data Analysis},
 52(12):5186--5201.

\bibitem[Hubert and Van Driessen, 2004]{Hubert:Discrim}
Hubert, M. and Van Driessen, K. (2004).
\newblock Fast and robust discriminant analysis.
\newblock {\em Computational Statistics \& Data Analysis},
45:301--320.

\bibitem[Koenker and Bassett, 1978]{Koenker:RegQuant}
Koenker, R. and Bassett, G. (1978).
\newblock Regression quantiles.
\newblock {\em Econometrica}, 46:33--50.

\bibitem[Lange et~al., 2014]{Lange:Classification}
Lange, T., Mosler, K., and Mozharovskyi, P. (2014).
\newblock Fast nonparametric classification based on data depth.
\newblock {\em Statistical Papers}, 55(1):49--69.

\bibitem[Li and Yu, 2008]{Li:SegmentationApproach}
Li, B. and Yu, Q. (2008).
\newblock Classification of functional data: A segmentation
          approach.
\newblock {\em Computational Statistics \& Data Analysis},
 52(10):4790 -- 4800.

\bibitem[Li et~al., 2012]{Li:DDplot}
Li, J., Cuesta-Albertos, J., and Liu, R. (2012).
\newblock {DD}-classifier: nonparametric classification
 procedure based on {DD}-plot.
\newblock {\em Journal of the American Statistical Association}, 
 107:737--753.

\bibitem[Liu, 1990]{Liu:Simplicial}
Liu, R. (1990).
\newblock On a notion of data depth based on random simplices.
\newblock {\em The Annals of Statistics}, 18(1):405--414.


\bibitem[L{\'o}pez-Pintado and Romo, 2006]{Lopez:ClassifDepth}
L{\'o}pez-Pintado, S. and Romo, J. (2006).
\newblock Depth-based classification for functional data.
\newblock In {\em Data depth: {R}obust {M}ultivariate {A}nalysis, 
  {C}omputational {G}eometry and {A}pplications}, volume~72 of 
	{\em DIMACS Ser. Discrete Math. Theoret. Comput. Sci.},
	pages 103--119. Amer. Math. Soc., Providence, RI.

\bibitem[Maronna et~al., 2006]{Maronna:RobStat}
Maronna, R., Martin, D., and Yohai, V. (2006).
\newblock {\em Robust Statistics: Theory and Methods}.
\newblock Wiley, New York.

\bibitem[Martin-Barragan et~al., 2014]{MartinBarragan:InterSVM_FDA}
Martin-Barragan, B., Lillo, R., and Romo, J. (2014).
\newblock Interpretable support vector machines for functional data.
\newblock {\em European Journal of Operational Research},
 232(1):146--155.

\bibitem[Mass\'{e} and Theodorescu, 1994]{Masse:Trimming}
Mass\'{e}, J.-C. and Theodorescu, R. (1994).
\newblock Halfplane trimming for bivariate distributions.
\newblock {\em Journal of Multivariate Analysis}, 48(2):188--202.

\bibitem[Mosler, 2013]{Mosler:DepthStat}
Mosler, K. (2013).
\newblock Depth statistics.
\newblock In Becker, C., Fried, R., and Kuhnt, S., editors,
 {\em Robustness and Complex Data Structures, Festschrift in
  Honour of Ursula Gather}, pages 17--34, Berlin, Springer.

\bibitem[Mosler and Mozharovskyi, 2016]{Mosler:DDClass}
Mosler, K. and Mozharovskyi, P. (2016).
\newblock Fast DD-classification of functional data.
\newblock {\em Statistical Papers}, doi: 10.1007/s00362-015-0738-3.

\bibitem[Nagy et~al., 2016]{Nagy:ConsistencyID}
Nagy, S., Gijbels, I., Omelka, M., and Hlubinka, D. (2016).
\newblock Integrated depth for functional data: statistical properties and consistency.
\newblock {\em ESAIM Probability and Statistics},  doi: 10.1051/ps/2016005.

\bibitem[Paindaveine and {\v{S}}iman, 2012]{Paindaveine:CompQuantiles}
Paindaveine, D. and {\v{S}}iman, M. (2012).
\newblock Computing multiple-output regression quantile regions.
\newblock {\em Computational Statistics \& Data Analysis},
 56:840--853.

\bibitem[Pigoli and Sangalli, 2012]{Pigoli:Wavelets}
Pigoli, D. and Sangalli, L. (2012).
\newblock Wavelets in functional data analysis: estimation of
 multidimensional curves and their derivatives.
\newblock {\em Computational Statistics \& Data Analysis},
 56(6):1482--1498.

\bibitem[Ramsay and Silverman, 2005]{Ramsay:BookFDA}
Ramsay, J. and Silverman, B. (2005).
\newblock {\em Functional Data Analysis}.
\newblock Springer, New York, 2nd edition.

\bibitem[Rossi and Villa, 2006]{Rossi:SVM_FDA}
Rossi, F. and Villa, N. (2006).
\newblock Support vector machine for functional data classification.
\newblock {\em Neurocomputing}, 69:730--742.

\bibitem[Rousseeuw and Hubert, 1999]{Rousseeuw:Regdepth}
Rousseeuw, P.~J. and Hubert, M. (1999).
\newblock Regression depth.
\newblock {\em Journal of the American Statistical Association},
 94:388--402.

\bibitem[Rousseeuw and Leroy, 1987]{Rousseeuw:RobReg}
Rousseeuw, P.~J. and Leroy, A. (1987).
\newblock {\em Robust Regression and Outlier Detection}.
\newblock Wiley-Interscience, New York.

\bibitem[Rousseeuw and Ruts, 1996]{Rousseeuw:Bivlocdepth}
Rousseeuw, P.~J. and Ruts, I. (1996).
\newblock Bivariate location depth.
\newblock {\em Applied Statistics}, 45:516--526.

\bibitem[Rousseeuw and Ruts, 1998]{Rousseeuw:Tukeymed}
Rousseeuw, P.~J. and Ruts, I. (1998).
\newblock Constructing the bivariate {T}ukey median.
\newblock {\em Statistica Sinica}, 8:827--839.

\bibitem[Rousseeuw and Ruts, 1999]{Rousseeuw:DepthPop}
Rousseeuw, P.~J. and Ruts, I. (1999).
\newblock The depth function of a population distribution.
\newblock {\em Metrika}, 49:213--244.

\bibitem[Rousseeuw et~al., 1999]{Rousseeuw:Bagplot}
Rousseeuw, P.~J., Ruts, I., and Tukey, J. (1999).
\newblock The bagplot: a bivariate boxplot.
\newblock {\em The American Statistician}, 53:382--387.

\bibitem[Rousseeuw and Struyf, 1998]{Rousseeuw:Ldepth}
Rousseeuw, P.~J. and Struyf, A. (1998).
\newblock Computing location depth and regression depth in
 higher dimensions.
\newblock {\em Statistics and Computing}, 8:193--203.


\bibitem[Ruts and Rousseeuw, 1996]{Ruts:isodepth}
Ruts, I. and Rousseeuw, P.~J. (1996).
\newblock Computing depth contours of bivariate point clouds.
\newblock {\em Computational Statistics \& Data Analysis},
 23:153--168.

\bibitem[Stahel, 1981]{Stahel:SDEst}
Stahel, W. (1981).
\newblock {\em Robuste Sch{\"a}tzungen: infinitesimale
 Optimalit{\"a}t und Sch{\"a}tzungen von Kovarianzmatrizen}.
\newblock PhD thesis, ETH Z{\"u}rich.


\bibitem[Struyf and Rousseeuw, 2000]{Struyf:DeepestLocation}
Struyf, A. and Rousseeuw, P.~J. (2000).
\newblock High-dimensional computation of the deepest location.
\newblock {\em Computational Statistics \& Data Analysis},
 34(4):415--426.

\bibitem[Thakoor and Gao, 2005]{Thakoor:Planes}
Thakoor, N. and Gao, J. (2005). 
\newblock Shape classifier based on generalized probabilistic 
descent method with hidden Markov descriptor. 
\newblock Tenth IEEE International Conference on Computer 
Vision (ICCV 2005), Vol. 1: 495--502. 

\bibitem[Tukey, 1975]{Tukey:depth}
Tukey, J. (1975).
\newblock Mathematics and the picturing of data.
\newblock In {\em Proceedings of the International Congress of
 Mathematicians}, Volume~2, pages 523--531, Vancouver.

\bibitem[Zuo, 2003]{Zuo:ProjDepth}
Zuo, Y. (2003).
\newblock Projection-based depth functions and associated medians.
\newblock {\em The Annals of Statistics}, 31(5):1460--1490.

\bibitem[Zuo and Serfling, 2000]{Zuo:Depth}
Zuo, Y. and Serfling, R. (2000).
\newblock General notions of statistical depth function.
\newblock {\em The Annals of Statistics}, 28:461--482.

\end{thebibliography}

\end{document}